\documentclass{aastex701}

\usepackage{amssymb}
\usepackage{csquotes}
\usepackage{epsfig}
\usepackage{mathtools}
\usepackage[version=4]{mhchem}
\usepackage{comment}
\usepackage{bm}
\usepackage{epstopdf}
\usepackage{amssymb}
\usepackage{geometry}
\usepackage[figuresright]{rotating}
\usepackage{placeins}
\usepackage{float}

\newcolumntype{H}{>{\setbox0=\hbox\bgroup}c<{\egroup}@{}}
\usepackage[caption=false]{subfig}
\renewcommand{\baselinestretch}{1.3}

\newcommand{\ditto}{$''$}

\newcommand{\kms}{\mbox{km~s$^{-1}$}}
\newcommand{\s}{\mbox{$''$}}

\newcommand{\my}{\mbox{$M_{\odot}$~yr$^{-1}$}}
\newcommand{\ls}{\mbox{$L_{\odot}$}}
\newcommand{\ms}{\mbox{$M_{\odot}$}}
\newcommand{\mdot}{$\dot{M}  $}

\newcommand{\tkin}{\mbox{$T_{\rm kin}$}}

\newcommand{\bfluxu}{\mbox{erg\,s$^{-1}$\,cm$^{-2}$}}

\newcommand{\moldens}{\mbox{$n_{H_2}$}}

\newcommand{\codos}{$^{12}$CO}
\newcommand{\cotres}{$^{13}$CO}
\newcommand{\dostotres}{$^{12}$C/$^{13}$C}
\newcommand{\cdos}{$^{12}$C}
\newcommand{\ctres}{$^{13}$C}
\newcommand{\coeksaat}{C$^{17}$O}

\newcommand{\oekchey}{$^{16}$O}
\newcommand{\oeksaat}{$^{17}$O}

\newcommand{\jocho}{J=8--7}
\newcommand{\jsaat}{J=7--6}
\newcommand{\jsix}{J=6--5}
\newcommand{\jcinco}{J=5--4}
\newcommand{\jchar}{J=4--3}
\newcommand{\jtres}{J=3--2}
\newcommand{\jdos}{J=2--1}
\newcommand{\junos}{J=1--0}
\newcommand{\hctresn}{H$^{13}$CN\,J=4--3}
\newcommand{\hctresnhi}{H$^{13}$CN\,J=8--7}

\newcommand{\hctresnmol}{H$^{13}$CN}
\newcommand{\hcnekpanch}{HC$^{15}$N}

\newcommand{\cssiete}{CS\,J=7-6}

\newcommand{\pscalcgs}{g\,cm\,s$^{-1}$}

\newcommand{\vdensunit}{cm$^{-3}$}

\newcommand{\engyunit}{erg}

\newcommand{\gsim}{\raisebox{-.4ex}{$\stackrel{>}{\scriptstyle\sim}$}}

\global\let\oldnewlabel\newlabel
\gdef\newlabel#1#2{\newlabelxx{#1}#2}
\gdef\newlabelxx#1#2#3#4#5#6{\oldnewlabel{#1}{{#2}{#3}}}
\let\newlabel\oldnewlabel
\begin{document}

\title{The Last Gasps of a Dying Star: ALMA Observations of the Pre-Planetary Nebula IRAS\,06530-0213}

\author{Raghvendra Sahai}

\affiliation{Jet Propulsion Laboratory, MS 183-900, California Institute of Technology, Pasadena, CA 91109, USA}
\email{raghvendra.sahai@jpl.nasa.gov}

\author{Wouter Vlemmings}
\affiliation{Department of Physics and Astronomy, Chalmers University of Technology, SE-412 96 Gothenburg, Sweden}
\email{wouter.vlemmings@chalmers.se}

\author{Guillermo Quintana-Lacaci}
\affiliation{Department of Molecular Astrophysics, Instituto de Física Fundamental (IFF-CSIC), C/ Serrano 123, 28006 Madrid, Spain}
\email{guillermo.q@csic.es}

\author{Hyosun Kim}
\affiliation{Korea Astronomy and Space Science Institute, 776 Daedeok-daero, Yuseong-gu, Daejeon 34055, Republic of Korea}
\email{hkim@kasi.re.kr}

\author{Javier Alcolea}
\affiliation{Observatorio Astronómico Nacional (IGN/CNIG), C/ Alfonso XII 3 y 5, E-28014 Madrid, Spain}
\email{j.alcolea@oan.es}

\author{Bruce Balick}
\affiliation{Department of Astronomy, University of Washington, Seattle, WA 98195, USA}
\email{balick@uw.edu}

\author{Eric G. Blackman}
\affiliation{Department of Physics and Astronomy, University of Rochester, Rochester, NY 14627}
\email{blackman@pas.rochester.edu}

\author{Arancha Castro-Carrizo}
\affiliation{ Institut de Radioastronomie Millimétrique, 300 rue de la Piscine, 38406 Saint-Martin-d’Hères, France}
\email{ccarrizo@iram.fr}

\author{Orsola de Marco}
\affiliation{Astrophysics and Space Technologies Research Centre, School of Mathematical and Physical Sciences, Macquarie University, Sydney, Australia}
\email{orsola.demarco@mq.edu.au}

\author{Joel H. Kastner}
\affiliation{Chester F. Carlson Center for Imaging Science, School of Physics \& Astronomy, and Laboratory for Multiwavelength Astrophysics, Rochester Institute of Technology, 54 Lomb Memorial Drive, Rochester, NY 14623, USA}
\email{jhkpci@rit.edu}

\author{Eric Lagadec}
\affiliation{Université Côte d’Azur, Observatoire de la Côte d’Azur, CNRS, Laboratoire Lagrange, France}
\email{elagadec@oca.eu}

\author{Chin-Fei Lee}
\affiliation{Academia Sinica Institute of Astronomy and Astrophysics, No.\,1, Sec.\,4, Roosevelt Road, Taipei 106216, Taiwan, R.O.C.}
\email{cflee@asiaa.sinica.edu.tw}

\author{Miguel Santander-Garc\'ia}
\affiliation{Observatorio Astronómico Nacional (OAN-IGN), Alfonso XII, 3, 28014, Madrid, Spain}
\email{m.santander@oan.es}

\author{Carmen S\'anchez Contreras}
\affiliation{Centro de Astrobiolog{'i}a (CAB), CSIC-INTA, ESAC Campus, Camino Bajo del Castillo s/n, E-28692 Villanueva de la Ca~nada, Madrid, Spain}
\email{csanchez@cab.inta-csic.es}

\author{Toshiya Ueta}

\affiliation{Department of Physics and Astronomy, University of Denver, Denver, CO 80208, USA}
\email{Toshiya.Ueta@du.edu}

\author{Albert A. Zijlstra}
\affiliation{Jodrell Bank Centre for Astrophysics, Department of Physics, The University iof Manchester, Manchester M13 9PL, UK}
\email{albert.zijlstra@manchester.ac.uk}

\author{Valentin Bujarrabal}
\affiliation{Observatorio Astron\'omico Nacional (OAN-IGN), Apartado 112, E-28803 Alcal\'a de Henares, Spain}
\email{v.bujarrabal@oan.es}

\author{Laurence Sabin}
\affiliation{Instituto de Astronomía, Universidad Nacional Autónoma de México, AP 106,  Ensenada 22800, BC, México}
\email{laurence.sabin@gmail.com}

\author{Daniel Tafoya}
\affiliation{Department of Physics and Astronomy, Chalmers University of Technology, Onsala Space Observatory, SE-439 92 Onsala, Sweden}
\email{daniel.tafoya@chalmers.se}

\begin{abstract}
We present high angular-resolution ($\sim0\farcs1-0\farcs5$) ALMA observations of millimeter-wave line and continuum emission (at $\sim0.44$ and $0.88$\,mm) in the pre-planetary nebula IRAS\,06530--0213 (IRAS\,06530). These data show the presence of an extended circular ring -- first evidence of the last thermal-pulse preceding the post-AGB phase in a carbon star -- and a central bipolar nebula (CBN) and torus.  The mass-loss rate of IRAS\,06530's AGB progenitor decreased immediately after the thermal pulse, then rose again just before the post-AGB phase as evidenced by the presence of filamentary arc structures around the CBN. The arc stuctures are likely part of a 3-D Archimidean spiral structure generally attributed to the presence of a binary companion. But we do not find a compact continuum source at the location of IRAS\,06530's central star, such as that associated with the compact dusty disks typically found in disk-prominent post-AGB objects known to have binary companions. The expansion ages derived for the torus and CBN imply that IRAS\,06530's progenitor transitioned to a post-AGB star $\lesssim700$\,yr ago. The molecular mass of the ejecta in IRAS\,06530 is dominated by the filamentary arc region with a mass $M_{H_2}=(0.12-0.25)$\,\ms. Compared with solar values, the torus of IRAS\,06530 appears to be significantly (modestly) enriched in $^{13}$C and \oeksaat~($^{15}$N) as well -- a pattern of rare-isotope enrichment inconsistent with standard nucleosynthesis models. From the luminosity of IRAS 06530 and the age of its detached shell, while it was still on the AGB, we infer that the mass of IRAS 06530's progenitor was $(1.6-3.4)$\,\ms.

\end{abstract}

\section{Introduction}\label{intro}
Understanding the impact of binary interactions on stellar evolution is a major challenge.  Binary star interactions dominate a substantial fraction of stellar phenomenology (\citealt{bond03, tylenda11, ivanova13}), and are believed to underlie the formation of most Planetary Nebulae (PNe), which represent the bright end-stage of most stars in the Universe. Such interactions are likely the key to the resolution of a long-standing puzzle: although PNe (and their progenitors, pre-PNe [PPNe]), evolve from slowly-expanding ($V_{exp}\sim$\,5-15\,\kms), generally round circumstellar envelopes (CSEs) of AGB stars, modern  surveys show that the vast majority of the former deviate strongly from spherical symmetry, showing a dazzling variety of bipolar, multipolar, and elliptical morphologies (e.g., \citealt{st98,ueta00,sahai07,siodmak08,sahai11,lagadec11}) and fast, collimated outflows ($V_{exp}\gsim$\,50-100\,\kms, e.g., \citealt{bujarrabal01,carmen12,imai07}). Dense, dusty, equatorial waists are also frequently present in PPNe and PNe, and recognized as an important morphological feature of this class of objects (e.g., \citealt{sahai07,sahai11}). About $(25-30)$\% of all known galactic post-AGB objects are known (or likely) binaries -- these have little or no outflows, but show prominent circumbinary disks (\citealt{winckel03,winckel25}), dubbed ``dpAGB" objects by \cite{setal11vla}.
Following intense debate over the past two decades (e.g., \citealt{balfrk02,apn3,demarco09,apn5,apn8e}), the current consensus is that the primary process for the formation and shapes of PNe occurs during the short-lived ($\sim1000$\,yr) PPN phase and the preceding very late-AGB phase (\citealt{st98,soker02}), and consists of hydrodynamic sculpting of the  progenitor AGB mass-loss envelopes, from the inside-out by wandering and/or episodic jets. This conclusion is strongly supported by estimates of the scalar momentum of PPN outflows from an analysis of CO and $^{13}$CO line profiles (mostly single-dish, but also interferometric mapping studies, typically with angular resolution  $>1\arcsec-2\arcsec$ (e.g., \citealt{cox2688,alcolea07,arancha10,sah08005}). The jets are believed to result from binary interaction, which can also produce equatorial density enhancements \citep{Decin20} commonly observed as dense waist structures in bipolar and multipolar PNe and PPNe (e.g., \cite{sahai07,sahai11}).

As a result of significant progress in the development of sophisticated simulations of jet launching and wind interaction (e.g., \citealt{lee03,bal13,bal19,bal20,staff16}), we now stand at the brink of a full understanding of the role of binary interactions in the demise of most stars in the Universe. Amongst the observational constraints necessary to motivate and guide the next generation of AGB-PN simulation tools aimed at understanding late-stage binary system evolution, an important one that is lacking is a measurement of masses and kinematics of the outflows and central regions (on scales small enough to resolve and isolate critical structures, e.g., base of outflow from central torus and/or disk) for a representative sample of PPNe.

We are therefore carrying out a comparative study of a representative sample of bipolar and multipolar post-AGB PPNe that have been imaged with HST and have beenpreviously detected in CO emission. The observational goal of this survey, dubbed PANORAMA ({\bf P}re-planet{\bf A}ry {\bf N}ebulae high-angular-res{\bf O}lution su{\bf R}vey) with {\bf A}l{\bf MA}), is to obtain high-fidelity maps of \codos\,\jtres~and \jsix~emission with $\sim0\farcs1$ resolution in order to probe the spatio-kinematic structure of the collimated outflows and the central disk/tori.

In this paper, we present ALMA observations of millimeter-wave CO lines in the bipolar PPN, IRAS\,06530--0213 (hereafter IRAS\,06530), and discuss their implications for our understanding of the near-circumstellar environment and mass-loss history of this fascinating pre-planetary nebula. This is a bipolar PPN, as revealed by optical HST imaging (\citealt{ueta00,sahai07}). The circumstellar chemistry is C-rich (C/O $=1.66\pm0.39$), its metallocity is sub-solar ([Fe/H]=$-0.21\pm0.11$ \citep{Desmedt16} and it shows the presence of $s$-process element enhancement (\citealt{Kamath22}), as well as the 21\,\micron~dust feature (\citealt{Kwok89}). Single-dish observations show \codos\,\jdos~emission, implying the presence of a molecular shell expanding at $\sim$13\,\kms~(\citealt{Hu94,Hrivnak05}). Determinations of the spectral type of the central star range between F5\,Ia (\citealt{Reddy96,Hrivnak03,Desmedt16}) and G1\,I (\citealt{Suarez06}). The optical spectrum shows molecular carbon absorption features (C$_2$ and C$_3$) (\citealt{Suarez06}). Photospheric abundance studies carried out using high-resolution spectra of IRAS\,06530 (\citealt{Winckel00,Reyniers04}) confirm the high carbon abundance, as well as the enhanced abundance of s-process elements, as expected for a post-AGB central star. Most recently, interferometric data of the \codos\,\jdos~emission  in IRAS\,06530 were obtained by \cite{sun25}, but with limited S/N and angular resolution to study in detail the gas distribution. The distance, $D$, to the central star of IRAS\,06530 is found to be $3.8$\,kpc, with a range of $(2.9-5.0)$\,kpc (\citealt{BailerJones21}) from its Gaia DR3 parallax $0.2412\pm0.0742$\,mas (\citealt{Lindegren21}).

The paper is organized as follows. In \S\,\ref{i06530obs}, we provide a summary of our 0.88 and 0.44 millimeter-wave observations of IRAS\,06530 used in this study. In \S\,\ref{struct}, we  describe the observational results, identifying the main structural components of IRAS\,06530 that include an extended ring around a central compact central bipolar nebula.

\S\,\ref{analysis} covers our analysis of these data as follows. In \S\,\ref{fbol-lum} we estimate the bolometric flux and luminosity, and in \S\,\ref{dustmodel}  (\S\,\ref{tot_mass}) we use the continuum (molecular-line) emission to derive the structure, temperature and mass of dust (gas) in IRAS\,06530, the gas-to-dust ratio, as well as abundance ratios of specific isotopologues of CO and HCN. In \S\,\ref{binary_arcs} we set constraints on a potential binary companion to IRAS\,06530's central star. In \S\,\ref{ages} we derive expansion ages and mass-loss rates for various structural components of IRAS\,06530. In \S\,\ref{discuss}, we discuss our results related to the spatio-kinematic structure of IRAS\,06530's and the isotope ratios that we find in it, in the context of theoretical models. We present our main conclusions in \S\,\ref{conclude}.

\section{Observations}\label{i06530obs}
We observed IRAS\,06530 with ALMA as part of program 2021.1.00182.S, in bands 7 \citep{Mahieu12} and 9 \citep{Baryshev15} in order to map the detailed structure of its mass ejecta. The phase center for our observations was located at R.A., Dec (J2000) = $06\,55\,31.820$, $-02\,17\,28.30$\footnote{all cordinates are in the ICRS system, epoch=J2000}. Data were obtained in two different ALMA configurations (C-3 and C-6 for band 7; C-3 and C-5 for band 9). Additional band 7 observations were also done using the ACA as part of (follow-up) program 2023.1.01128.S. The band 7 C-6, C-3 and ACA data were taken on 2022-07-24, 2022-09-16, and 2023-10-04, respectively; the band 9 C-5 and C-3 data were taken on 2024-09-28 and 2024-08-10, respectively.

We obtained the calibrated data sets from the archive. A summary of the molecular line discussed in this paper is listed in Table\,\ref{tbl-obslog}. The names of dataset for band 7 (band 9) C-6 \& C-3 (C-5 \& C-3) incorporate TM1 (observations with relatively long baselines) and TM2 (observations with shorter baselines) in them; we will hereafter refer to these as TM1 and TM2 data. We produced datasets combining the different configurations during image reconstruction using {\it tclean} in CASA 6.6.1, these are referred to as TM1+TM2 and TM1+TM2+ACA. We have used the combined images using the TM1+TM2 datasets for most of our analysis as these configurations are common between the two bands.

In order to produce the continuum images and perform continuum subtraction using the casa task {\it uvcontsub}, we identified line emission regions and regions with increased noise (likely due to atmospheric lines) by investigating the visibility spectra. Continuum images were made of the line free channels, using natural weighting and multi-scale clean, combining all different configurations. The emission lines cubes were created using all data, as well as (for band 7) excluding the ACA. In all cases we used multi-scale clean and for most lines except the weakest ones we used Briggs weighting with a robust parameter of 0.5\footnote{https://casa.nrao.edu/UserMan/UserMansu263.html}. Unless otherwise noted, the cubes were created with 150 channels, each of width 0.5\,\kms, in the range $V_{\rm lsr}=-7$\,\kms~to $67$\,\kms\footnote{All velocities are referred to the LSRK frame}.

\section{The Structure of IRAS\,06530}\label{struct}
In Band 7, we detected emission from the \jtres~lines of \codos~and \cotres, the \jchar~lines of \hctresnmol~and \hcnekpanch\, and the \jsaat~line of CS. In Band 9, the \jsix~lines of \codos~and \coeksaat, and the \jocho~line of \hctresnmol~were detected.

The TM1+TM2+ACA map of the \codos\,\jtres~line at the systemic velocity (V$_{lsr}=30.6$\,\kms) shows a bright central elongated source (size $\sim 2\farcs5\times1\farcs3$) surrounded by a fainter circular ring (diameter$\sim9\arcsec$) (Fig.\,\ref{i06530_co32_tm1_tm2}). Filamentary arc structures can be seen around the immediate periphery of the elongated source.

Moment-0 maps using broad velocity-ranges redwards, bluewards, and centered on the systemic velocity show the presence of very faint structured emission between the region showing the arcs and the ring (Fig.\,\ref{i06530_co32_azimuthal}), which is not azimuthally symmetric, but enhanced in specific azimuthal and radial regions. The intensities in the brightest of these enhancements typically have a signal-to-noise ratio of $S/N=3-4$. The average intensity in the NNW azimuthally bright region (i.e., in the range $PA=-25$\arcdeg\,to\,$30\arcdeg$~and radius=$(2\farcs5-3\farcs3)$) in the central 30\,\kms~around the systemic velocity (Fig.\,\ref{i06530_co32_azimuthal}c) is a factor 2.5 higher compared to the noise in the diametrically-opposed region with no detectable emission.

The TM1+TM2+ACA \codos\,\jtres~channel maps (Fig.\,\ref{i06530_co32_tm2_ch}) show that both the ring and central source have their maximum angular extent at velocities near the line-center and become progressively smaller towards the line wings, implying that both of these are radially-expanding structures. The ring is thus a part of a geometrically-thin, expanding spherical shell. The bright central source lies within an ellipse of size $\sim3\farcs4\times3\farcs0$. A spectrum extracted from this region (``bipolar central-source") using the TM1 \codos\,\jtres~datacube also shows a centrally-peaked profile (hereafter ``central-source spectrum" -- Fig.\,\ref{co32spec}, green curve); with prominent local emission peaks at blue-(red-) shifted velocities that arise in the front and back parts of the ring structure lying along the line-of-sight ($los$) towards the central source.

A spatially-integrated \codos\,\jtres~spectrum of the whole nebula, extracted from the TM1 data\footnote{we have used the TM1 data here as it best separates the various structural components of IRAS\,06530} using a circular aperture of diameter $11\farcs5$, shows a centrally-peaked, parabolic-shaped profile atop a wider plateau (Fig.\,\ref{co32spec}, black curve); the plateau emission is more prominently visible near the blue-shifted edge of the profile. A spectrum of the central bipolar nebula using an elliptical aperture of size $\sim3\farcs4\times1\farcs5$ (green curve) shows a similar parabolic-shaped profile (but more dominant) together with local peaks in the wings. The latter are due to emission from the radially-expanding front and back parts of the ring projected against the central source; these peaks are centered at 18.2 and 43\,\kms~(V$_{lsr}$), implying a ring expansion velocity\footnote{the term ``expansion velocity", here and elsewhere, means the velocity of the outflowing material relative to the systemic velocity} of V$_{e(rng)}=12.4$\,\kms~and a systemic velocity of V$_{lsr}$=30.6\,\kms -- the latter is consistent with the stellar radial velocity listed in GAIA DR3, $V_{hel}=47.21\pm1.43$ or V$_{lsr}=29.9\pm1.43$\,\kms. A fit to the central parabolic component, using the circumstellar line-shape function from \cite{wannier86}, with $\alpha>>1$, gives a (projected) expansion velocity of the bipolar central source of $V_{e(bip)}\sim11$\,\kms, however, given its non-spherical morphology, the actual expansion velocity is likely higher (see \S\,\ref{cbn_desc}).

\subsection{The Circular Ring}\label{ring_desc}

A radial cut of the \codos\,\jtres~TM1+TM2 image intensity at the systemic velocity, from the central star to radii larger than the ring and averaged over all azimuthal angles, shows a bright central peak due to the bright central source, and a weaker peak farther out due to the ring (Fig.\,\ref{i06530_rcut_co32_tm1}a). The ring emission consists of a sharp component (hereafter Ring Sharp component, or R-Shrp) atop a broader lower-intensity plateau (hereafter Ring Plateau component, or R-Plat) (Fig.\,\ref{i06530_rcut_co32_tm1}b, black curve).
We made two-Gaussian fits to the ring emission intensity.

We find that R-Shrp is centred at $r_{shrp}=4\farcs3\pm0\farcs001$ (16,375\,au), and R-Plat is centered at $r_{plat}=4\farcs64\pm0\farcs008$ (17,515\,au); the peak intensity of R-Shrp, $4$\,K, is a factor $1.8$ higher than that of R-Plat (Fig.\,\ref{i06530_rcut_co32_tm1}b). The observed width (FWHM) of the R-Shrp (R-Plat) component is $0\farcs30\pm0\farcs004$ ($0\farcs80\pm0\farcs01$).

In most of the region between the radial intensity peaks due to the central bipolar source and the ring, i.e., in the $\sim1\farcs5-3\farcs5$ range (hereafter ``cavity"), the average intensity at systemic velocity, is negligible.

The ring is not seen directly in the \cotres\,\jtres~TM1+TM2 channel maps due to inadequate sensitivity; we therefore averaged the central 15 channels (i.e., 7.5\,\kms) of the datacube around the line-center and extracted a radial intensity cut from the resulting image (Fig.\,\ref{i06530_rcut_13co32_tm1tm2}). The channel range used for the averaging is motivated by the fact that in this velocity range, the dimensions of the ring do not change much as seen in the \codos\,\jtres~TM1+TM2 channel maps (the ring radius at $\pm\sim4$\,\kms~from line-center is 97.5\% of that at the line-center, as derived from the \codos\,\jtres~data).  The intensity cut also shows the presence of the R-Shrp component --
a Gaussian fit shows that it is centred at $\sim4\farcs3\pm0\farcs007$, with a width (FWHM) of $\sim0\farcs33\pm0\farcs015$, as in the case of \codos\,\jtres. However the presence of a negative dip in the intensity at smaller radii implies that the R-Shrp component's peak intensity and width cannot be characterized reliably.

The ring is not seen in Band 9 data. It is seen faintly in the (much more noisy) \codos\,\jdos~imaging by \cite{sun25}, but its structure is poorly defined; e.g., these authors state that the ring is imperfect and rather than a circular shape, it has a Q-shape, however we do not find any evidence of such a shape. We believe their inference about the shape comes from an analysis of emission artifacts seen in contours at low significance levels.

\subsection{The Central Bipolar Nebula (CBN) and Filamentary Arc Region (FAR)}\label{cbn_desc}
\begin{figure}[htbp]

\includegraphics[width=11cm]{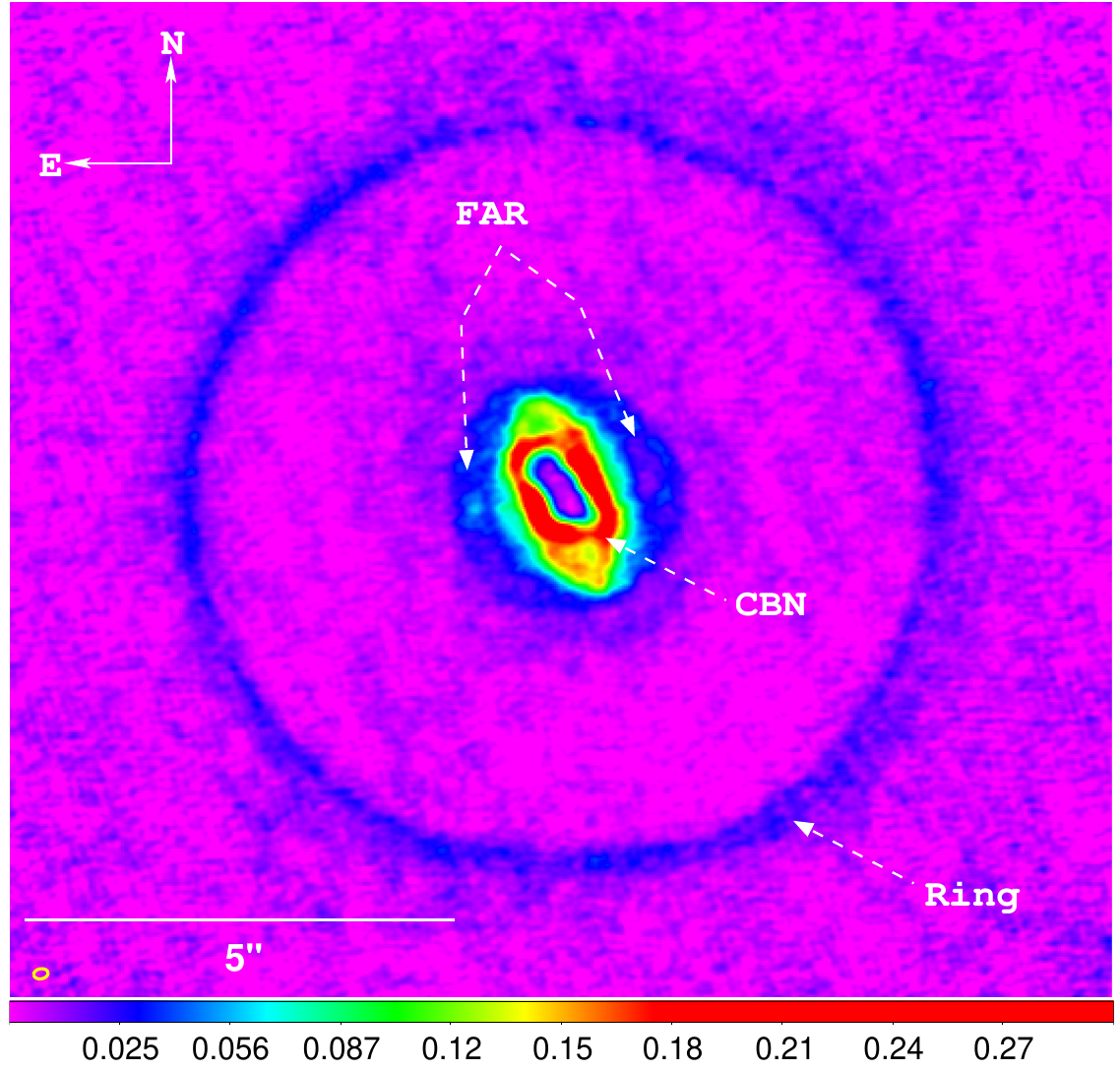} 

\caption{The TM1+TM2+ACA map of the \codos\,\jtres~emission towards IRAS\,06530 at the systemic velocity (averaged over a velocity width of $2$\,\kms), showing its different substructures -- the ring, the central bipolar nebula (CBN), and the filamentary arc region (FAR). Beam is $0\farcs17\times0\farcs12$, $PA=-80.14$\arcdeg. Scale bar shows intensity in units of Jy\,beam$^{-1}$.
}
\label{i06530_co32_tm1_tm2}
\end{figure}

\begin{figure}[htbp]
\includegraphics[width=16cm]{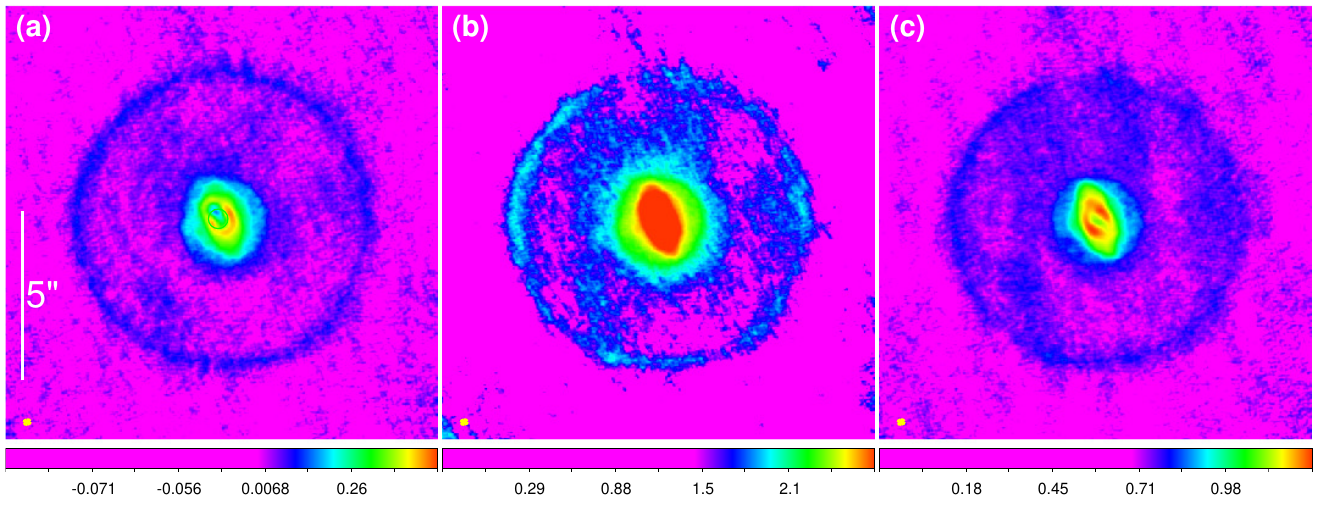}
\caption{The TM1+TM2+ACA moment 0 maps of the \codos\,\jtres~emission towards IRAS\,06530 showing the very faint azimuthally-asymmetric emission between the FAR and the ring in different velocity ranges. The integrated intensity 10\,\kms~immediately bluewards (redwards) of the systemic velocity is shown in panel $a$ ($c$), whereas panel $b$ shows the integrated intensity over the central 30\,\kms~of the line. Beam is $0\farcs17\times0\farcs12$, $PA=-80.14$\arcdeg, shown as yellow ellipse. Scale bars show intensity in units of Jy\,beam$^{-1}$. North (East) is up (to the left).}
\label{i06530_co32_azimuthal}
\end{figure}

\begin{figure}[htbp]

\includegraphics[width=16cm]{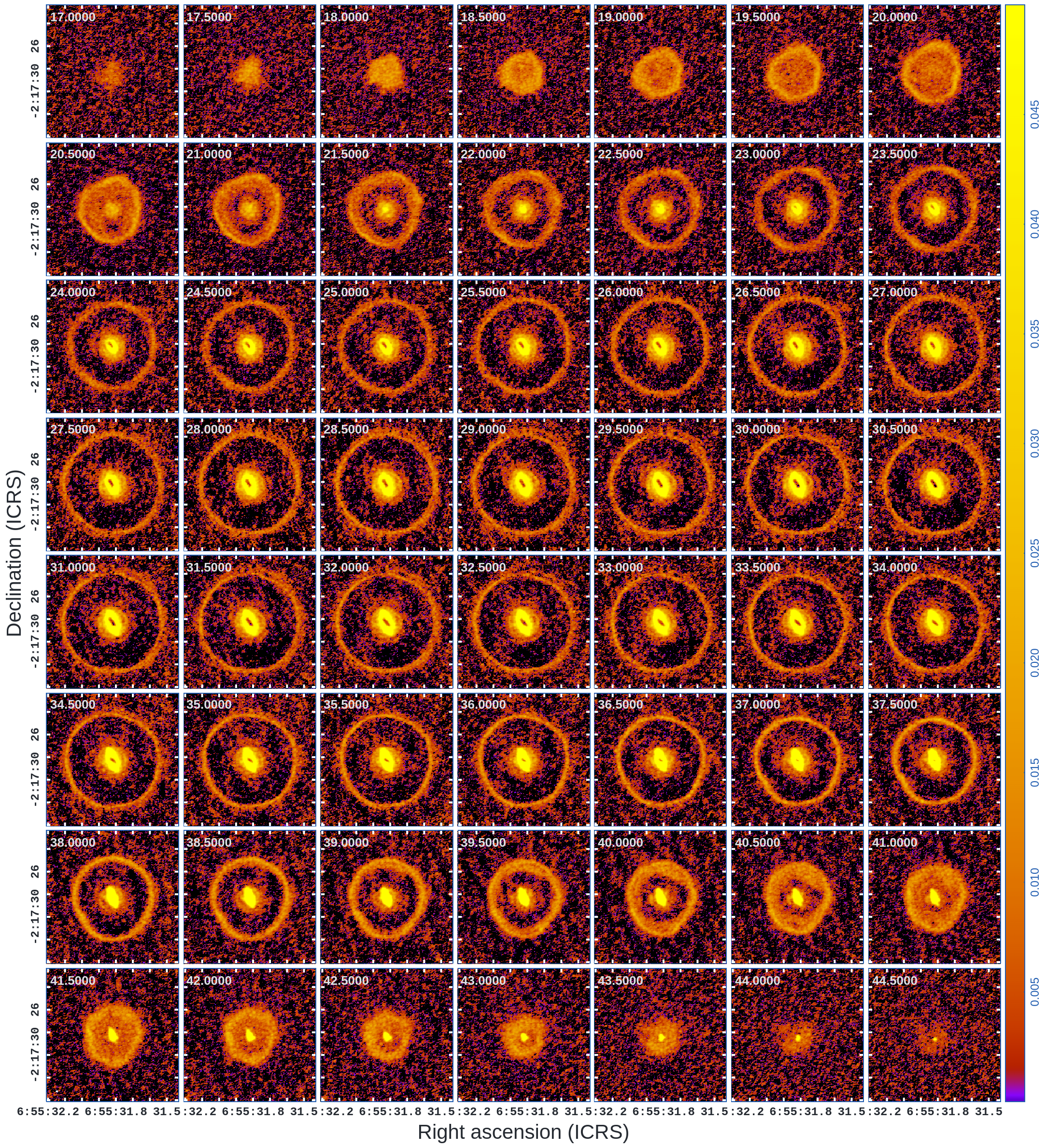}
\caption{Channel maps from the combined TM1+TM2+ACA datacube of the \codos\,\jtres~emission towards IRAS\,06530 showing the spatio-kinematic structure of the ring. Beam is $0\farcs17\times0\farcs12$, $PA=-80.14$\arcdeg. Scale bar shows intensity in units of Jy\,beam$^{-1}$.}
\label{i06530_co32_tm2_ch}
\end{figure}

The bright central source shows two components in \codos\,\jtres~emission at the systemic velocity -- (1) a brighter one which is bipolar with a central elongated cavity (hereafter central bipolar nebula or ``CBN"), and (2) a fainter one that consists of filamentary arc structures atop diffuse emission (hereafter Filamentary Arc Region, or ``FAR") (Fig.\,\ref{i06530_co32_tm1_tm2}, with radial extents spanning $\sim1\farcs1-1\farcs4$, that appear to emanate from the tips of the bipolar structure along its long axis (Fig.\,\ref{i06530_bipolar}a).

The shape of the CBN is point-symmetric in detail.  Two dashed lines mark two symmetry axes passing through the central star in Fig.\,\ref{i06530_bipolar}. ``Axis 1" (${\rm PA}=22\arcdeg\pm2\arcdeg$) aligns with the NE-SW tips of the bipolar shape as seen in \codos\,\jtres, \codos\,\jsix, and \cotres\jtres ~(Fig.\,\ref{i06530_bipolar}a,b,c) emission at the systemic velocity and corresponds to the axis of the larger outer lobes visible in the HST 0.55\,\micron~(filter F555W) image (Fig.\,\ref{i06530_bipolar}f). ``Axis 2" (${\rm PA}=42\arcdeg\pm5\arcdeg$) follows the roughly trapezoidal inner periphery defining the central cavity, derived from the \codos\,\jsix~(TM1+TM2) image at line-center, where the inner lobes predominantly emit. Axis 2 also corresponds to that of the smaller pair of lobes (inner lobes) seen in the HST image (Fig.\,\ref{i06530_bipolar}f). A third line (with ${\rm PA}=112\arcdeg\pm2\arcdeg$, i.e., perpendicular to Axis 1) joins two local peaks seen in CBN's waist in the TM1 \hctresn~emission at the systemic velocity (Fig.\,\ref{i06530_bipolar}d) that presumably define the cross-section of a dense torus in the CBN. \hctresn~and \cssiete~were specifically included in the spectral windows used for the observations in order to trace high-density regions; \cssiete~is fainter but shows roughly the same structure as \hctresn. Faint (and noisy) emission from \hcnekpanch, largely confined to the torus region, can be seen in a moment 0 map made using the TM2 data (not shown). Continuum emission at 0.89\,mm is shown in (Fig.\,\ref{i06530_bipolar}e) (discussed in detail in \S\,\ref{contdata}). Moment 0 (TM1) maps \cssiete, \hctresn, and \cotres\,\jtres~also show faint (but noisy) emission extended along the major axis of the nebula, presumably arising from the bipolar lobes (Fig.\,\ref{i06530_13co32_h13cn_cs76_mom0}).

\begin{figure}[htbp]

\includegraphics[width=16cm]{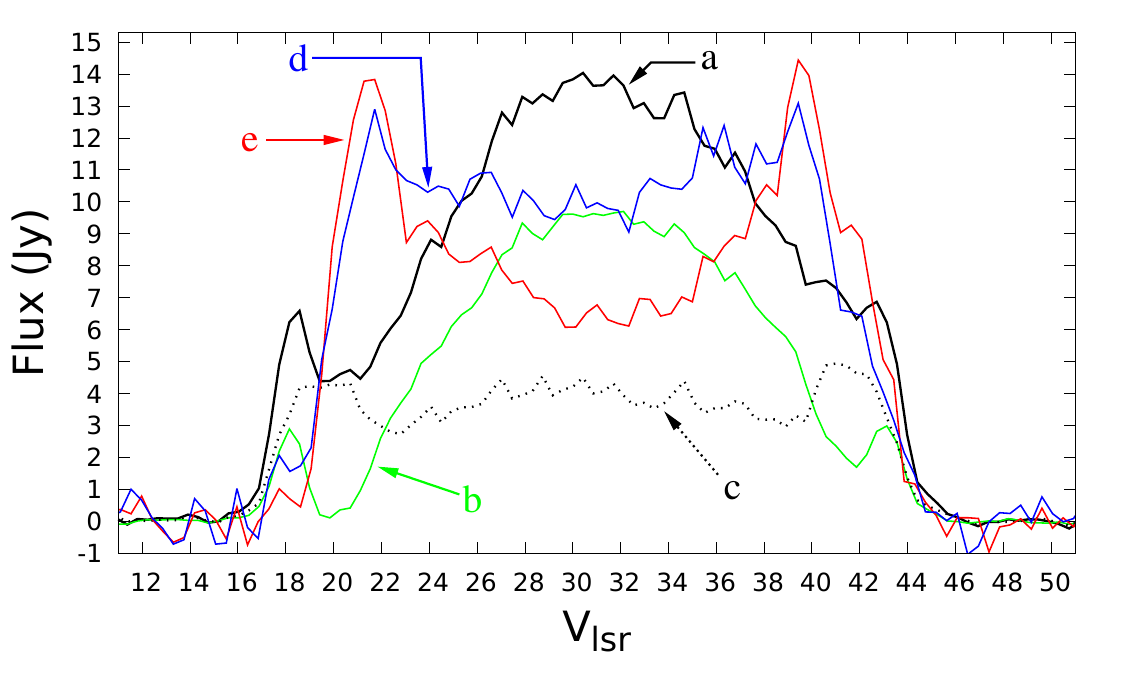}

\caption{Spatially-integrated \codos\,and \cotres\,\jtres~spectra of IRAS\,06530, extracted using different apertures: \codos\,\jtres~spectrum of (a) full nebula using circular aperture of diameter $11\farcs5$ (black curve), (b) the central bipolar nebula using an elliptical aperture of size $\sim3\farcs4\times1\farcs5$ (green curve), (c) their difference (black dashed curve) (using the TM1 datacube); \cotres\,\jtres~spectra of the (d) central bipolar nebula (blue curve) and (e) torus (red curve), respectively, (using the TM2 datacube, and scaled up by factors of $30$ and $100$).}
\label{co32spec}
\end{figure}

\begin{figure}[htbp]

\rotatebox{0}{\includegraphics[width=15.4cm]{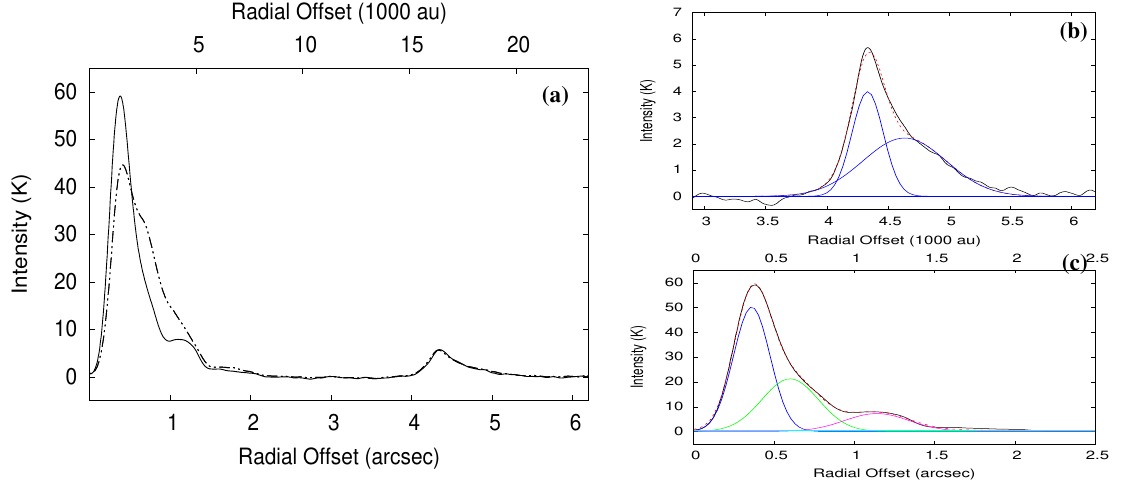}}

\vskip 0.01in
\caption{Radial intensity cuts of the \codos\,\jtres~emission at systemic velocity (TM1+TM2) (central star is at r=0): (a) averaged over (i) all azimuthal angles (dash-dot-dot curve), (ii) azimuthal wedges covering the position angle ranges $-107\arcdeg \le PA \le -35\arcdeg$ and $73\arcdeg \le PA \le 145\arcdeg$ (solid curve), (b) two-Gaussian fit (total = dashed red curve, R-shrp and R-Plat components = blue curves) to the ring emission (black curve) in a\,(i), together with a flat linear component, and (c) 3-Gaussian fit to the radial intensity in a\,(ii) (black curve) within the central region of IRAS\,06530 (total = dashed red curve); the innermost two components (blue and green curves) define the CBN emission, whereas the third, weakest component (magenta curve) defines the FAR emission, together with a flat linear component.}
\label{i06530_rcut_co32_tm1}
\end{figure}

\begin{figure}[htbp]

\rotatebox{0}{\includegraphics[width=12cm]{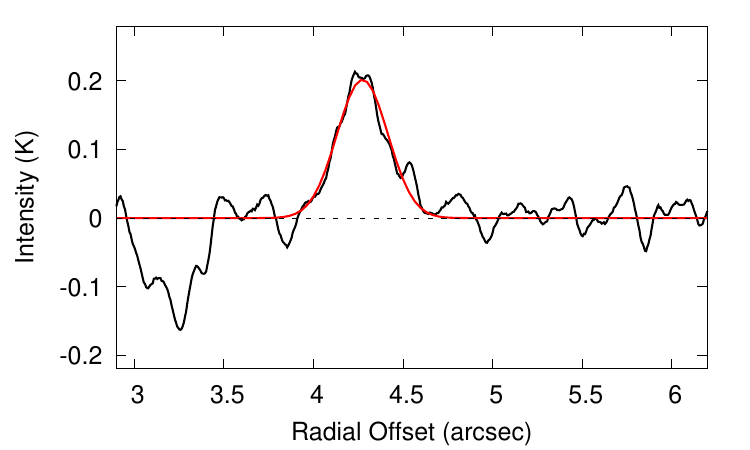}}
\vskip 0.01in
\caption{Radial intensity cut of the \cotres\,\jtres~emission (TM1) (central star is at r=0), averaged over all azimuthal angles (black curve). Also shown, is a Gaussian fit to the ring emission (red curve).}

\label{i06530_rcut_13co32_tm1tm2}
\end{figure}

\begin{figure}[htbp]

\includegraphics[width=16cm]{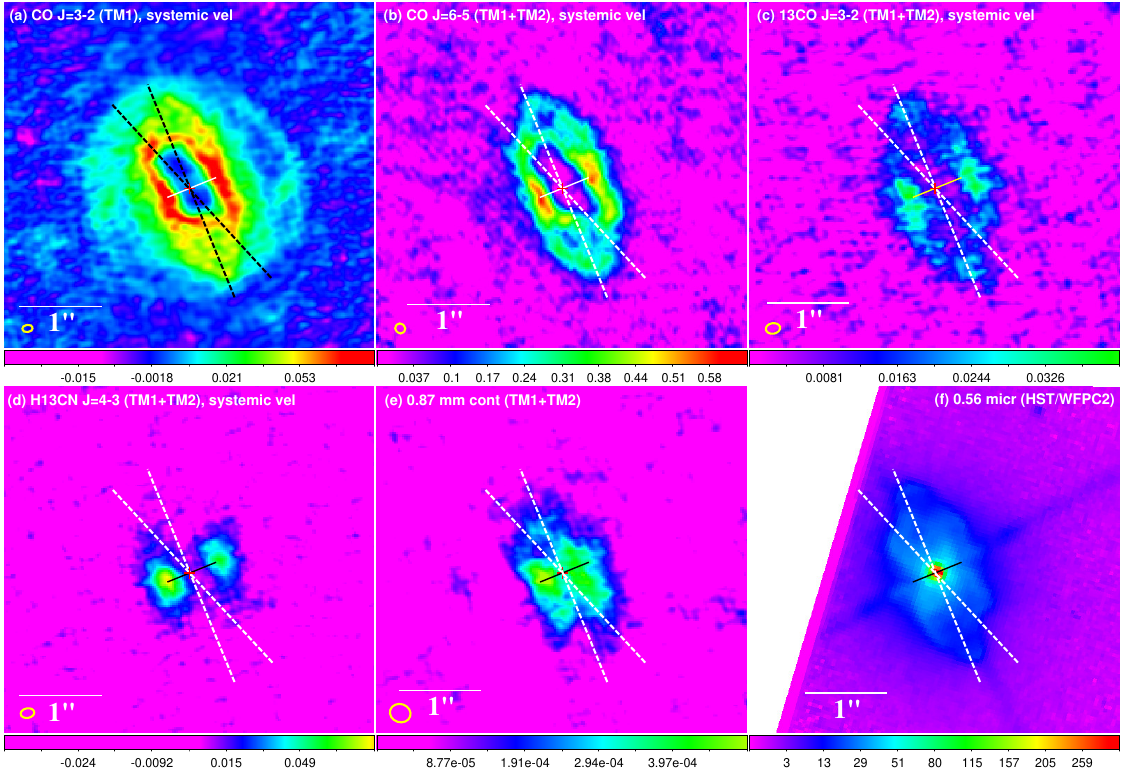}  
\caption{The central bipolar nebula (CBN) in IRAS\,06530: (a) \codos\,\jtres~emission  (TM1) at the systemic velocity (averaged over a velocity width of 1.27\,\kms), (b) \codos\,\jsix~emission  (TM1+TM2) at the systemic velocity (averaged over a velocity width of 1.5\,\kms), (c) \cotres\,\jtres~emission  (TM1+TM2) at the systemic velocity (averaged over a velocity width of 1.5\,\kms), (d) \hctresn~emission (TM1+TM2) at the systemic velocity (averaged over a velocity width of 2.5\,\kms), (e)  0.89\,\micron~continuum (TM1+TM2), (f) HST/WFPC2 F555W image (processed to enhance sharp structures as in Fig.\,1, Sahai \& Trauger 1998). Crosses show the location of the central star; long dashed white lines show two symmetry axes (with position angles ${\rm PA}=22\arcdeg$ and ${\rm PA}=38\arcdeg$) that go through central star and characterise the point-symmetric shape of the nebula; small dashed line (${\rm PA}=112\arcdeg$) joins two local peaks in the \hctresn~image that define the cross-section of the toroidal nebular torus. Beams are (a) $0''.13\times0''.090$, $PA=-82.6\arcdeg$, (b) $0''.13\times0''.12$, $PA=47.3\arcdeg$, (c) $0''.18\times0''.13$, $PA=-80.8\arcdeg$, (d) $0''.17\times0''.12$, $PA=-80.1\arcdeg$, and (e) $0''.26\times0''.22$, $PA=161.1\arcdeg$, and shown as yellow ellipses in the panels. The PSF in the HST image in panel $f$ is $0\farcs06$. Scale bars in panels $a-e$ show intensity in units of Jy\,beam$^{-1}$.
}

\label{i06530_bipolar}
\end{figure}

\begin{figure}[htbp]
\includegraphics[width=16cm]{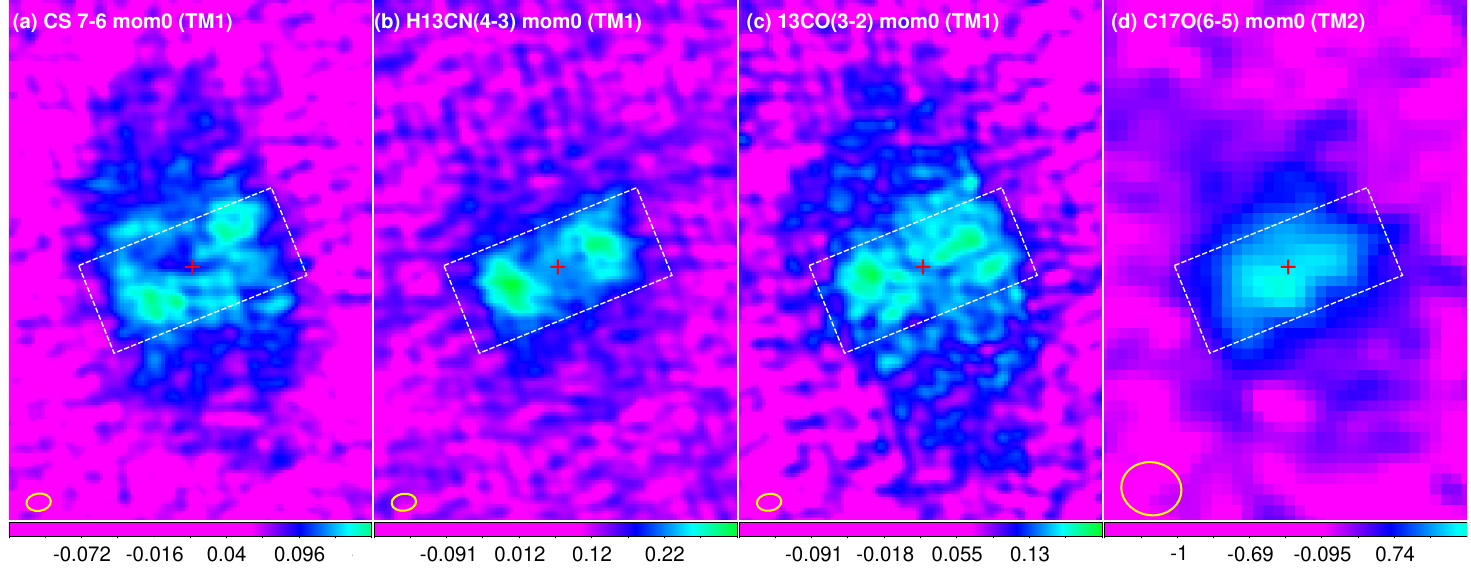}
\caption{Moment 0 maps of $(a-c)$ \cssiete, \hctresn, and \cotres\,\jtres~(TM1) showing bright emission from the torus region, as well as faint (but noisy) emission extended along the major axis of the nebula, arising from the bipolar lobes, ($d$) \coeksaat\,\jsix~(TM2) moment 0 map showing faint emission from the torus region. Cross shows central star location at RA=6:55:31.821, Dec=-2:17:28.24, and rectangle with sides $1\farcs1\times0\farcs5$, ${\rm PA}=22$\arcdeg~shows the aperture used to extract the dust continuum emission from the torus. Beams are (a) $0''.13\times0''.092$, $PA=-82.4\arcdeg$, (b) $0''.13\times0''.090$, $PA=-82.6\arcdeg$, (c) $0''.13\times0''.09$, $PA=-82.6\arcdeg$, and (d)  $0''.32\times0''.28$, $PA=78.3\arcdeg$, and shown as yellow ellipses in the panels. Scale bars show intensity in units of Jy\,beam$^{-1}$.
}
\label{i06530_13co32_h13cn_cs76_mom0}
\end{figure}

A spectrum extracted from an ellipse of size $\sim2\farcs4\times1\farcs4$ \footnote{using the TM2 datacube because of its higher sensitivity to the relatively weaker \cotres~emission} that fully encompasses the \cotres\,\jtres~emission shows a double-peaked profile with relatively steep line wings (Fig.\,\ref{co32spec}, blue curve).
A similary extracted \cotres\,\jtres~spectrum of just the waist region (using a rectangular \enquote{waist-aperture}, see \S\,\ref{contdata} below) also shows a double-peaked profile with the blue- and red-shifted peaks located at V$_{lsr}=21.5$ and $39.5$\,\kms~(Fig.\,\ref{co32spec}, red curve), confirming the systemic velocity of V$_{lsr}=30.6$\,\kms~(as derived above), and a waist expansion velocity of $V_{e(wst)}\sim9$\,\kms, which is less than $V_{e(bip)}=11$\,\kms~and implies that the waist region is expanding slower than the rest of the bipolar nebula. The sloping wings of the \cotres\,\jtres~profile beyond each of the peaks indicates substantial turbulence in the emitting material, $\delta V_{turb}\sim2$\,\kms, estimated from the average of the difference of the velocity at the peak and at half the peak-intensity for each of the blue- and red-shifted peaks of the bipolar source profile (1.9 and 2.1\,\kms).

In order to estimate the typical expansion velocity of the lobe walls, we extracted a spectral profile of the lobe region, using a thin rectangular aperture centered at a representative location offset by $0\farcs44$ along axis-1 from the center of the torus, with its long side perpendicular to axis-1, using the \cotres\,\jtres~TM1 datacube. The profile (not shown) shows a velocity-width that has the same width (at zero-intensity) as that of one extracted from the torus region -- assuming that the material in the lobe-walls is expanding radially and the CBN's long axis is within $10$\arcdeg~of plane of the sky, we find that the total (3-D) expansion velocity in the lobe at this location is $\sim16.5$\,\kms.

\begin{figure}[htbp]
\includegraphics[width=16cm]{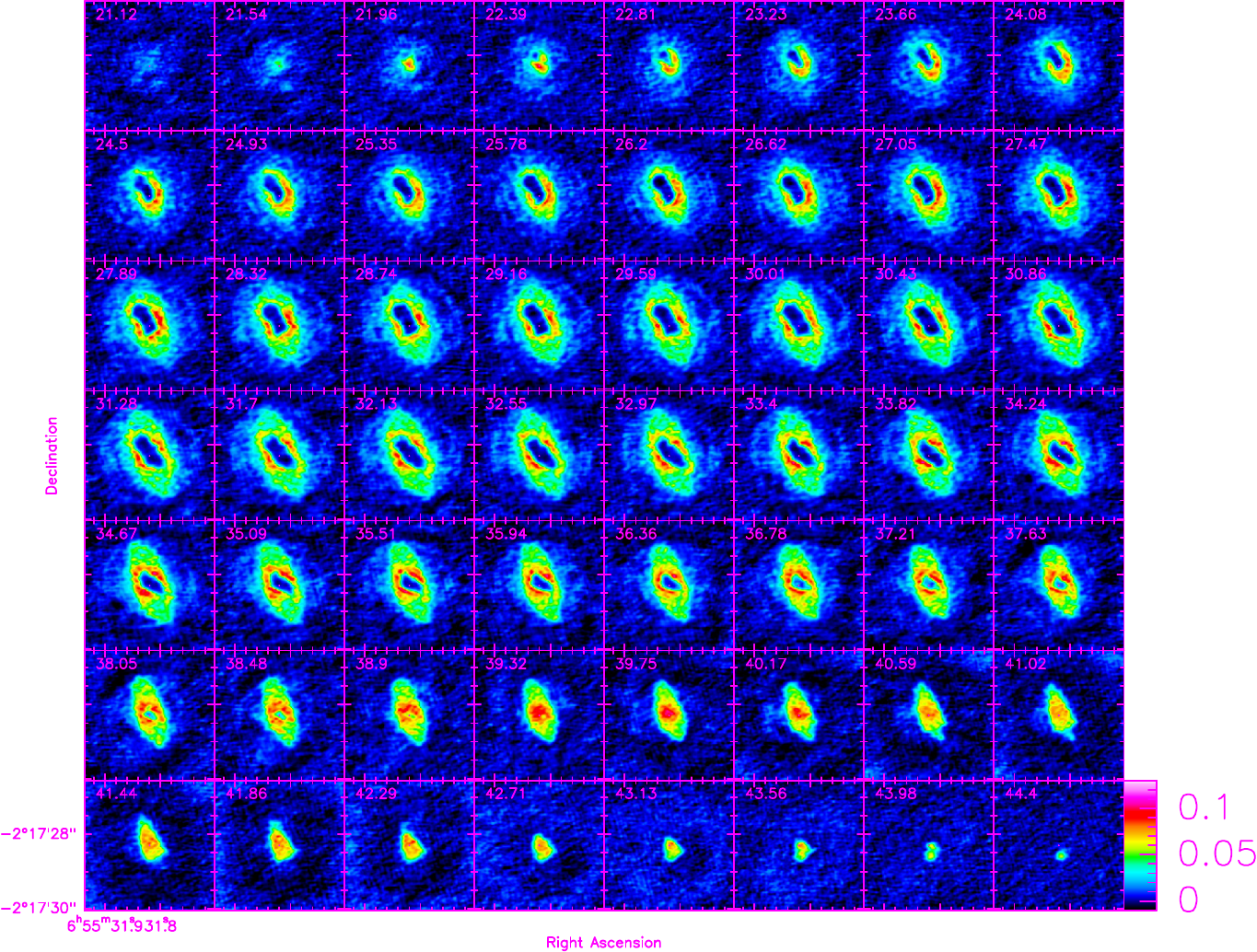}
\caption{Channel maps of the \codos\,\jtres~emission (TM1) towards IRAS\,06530 showing the central bipolar structure. Beam is $0''.13\times0''.09$, $PA=-82.6\arcdeg$. Scale bar shows intensity in units of Jy\,beam$^{-1}$.}
\label{i06530_co32_tm1_ch}
\end{figure}

The \codos\,\jtres~channel maps of the central nebula (Fig.\,\ref{i06530_co32_tm1_ch}) show that for velocities that are progressively more blue (red) -shifted from the central velocity, (i) the central cavity becomes smaller, and its center moves progressively towards the NE (SW). 

\begin{figure}[htbp]

\includegraphics[width=16cm]{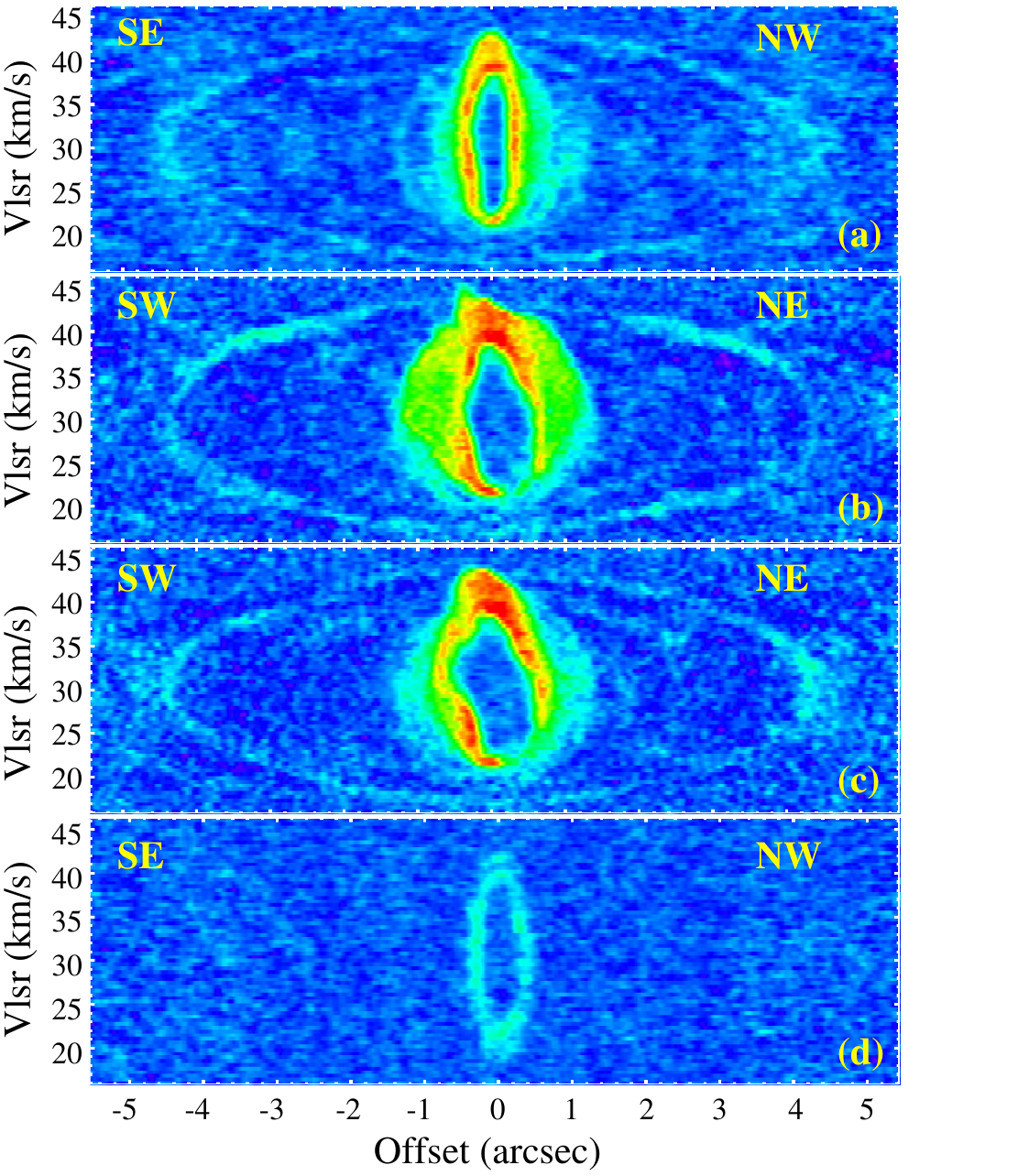}
\caption{Position-velocity cuts extracted from the \codos\,\jtres~emission (TM1) datacube, taken along the (a) torus (i.e., ${\rm PA}=112$\arcdeg), (b) axis 1 (i.e., ${\rm PA}=22$\arcdeg), (c) axis 2 (i.e., ${\rm PA}=42$\arcdeg). Panel (d) shows the torus PV cut in \hctresn. The cuts are $0\farcs09$-wide and pass through the location of the central star.}
\label{i06530_co32_tm1_pv}
\end{figure}

\begin{figure}[htbp]
\rotatebox{270}{\includegraphics[width=8cm]{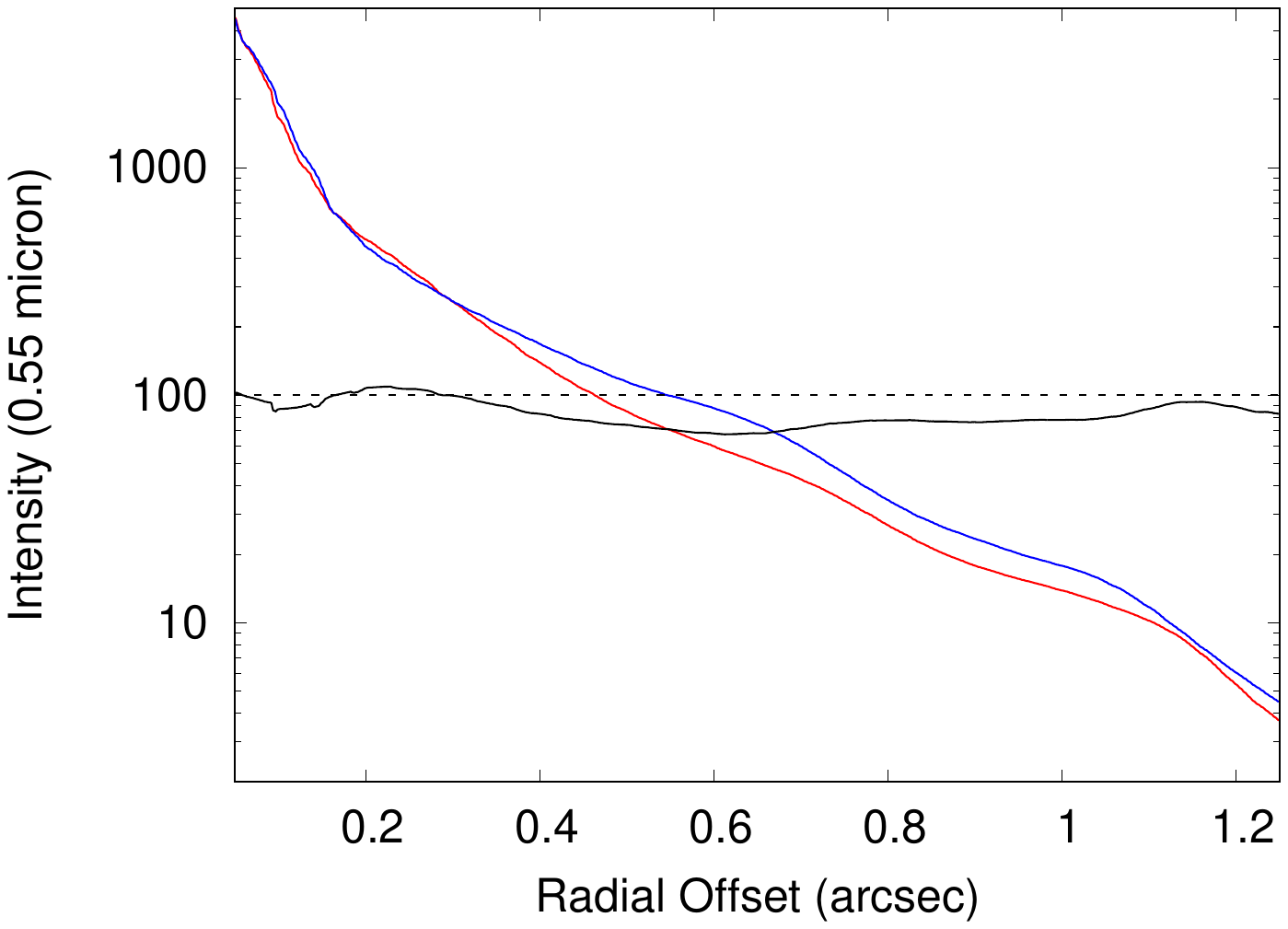}}
\vskip 0.5in
\caption{The 0.55\,\micron~radial intensity of the lobe structure to the NE (blue curve) and to the SW (red curve) of the central star (using 60\arcdeg~wide angular wedges around axis 1, and their ratio (shown as a percentage) (SW/NE, black curve).}
\label{i06530_f555w-rcut}
\end{figure}

Position-velocity (PV) cuts along the waist and axis clearly reveal the expanding nature of the CBN, as well as the ring and the faint filamentary arc structures that lie between the bipolar nebula and the ring (Fig.\,\ref{i06530_co32_tm1_pv}). The PV cut along the waist (Fig.\,\ref{i06530_co32_tm1_pv}a) shows a (roughly) limb-brightened elliptical shape with no obvious velocity gradient, as expected for an expanding torus seen edge on. PV cuts along axis 1 and 2 (Fig.\,\ref{i06530_co32_tm1_pv}b,c) also show a (roughly) limb-brightened  elliptical shape but with a small velocity gradient, such that the NE (SW) tip is blue (red)-shifted, implying that the outflows along axis 1 and 2 are slightly inclined to the sky-plane, with the NE side being closer to us. The 0.55\,\micron~radial intensity (extracted from the HST F555W image) of the lobe structure to the NE and to the SW of the central star (extracted using 55\arcdeg~wide angular wedges) (Fig.\,\ref{i06530_f555w-rcut}) shows that the SW lobe is slightly fainter than the NE one (by $\sim10-20$\,\%), as expected if the NE lobe is closer to us and therefore the light from it suffers somewhat less extinction by the ambient circumstellar medium.

\begin{figure}[htbp]
\includegraphics[width=16cm]{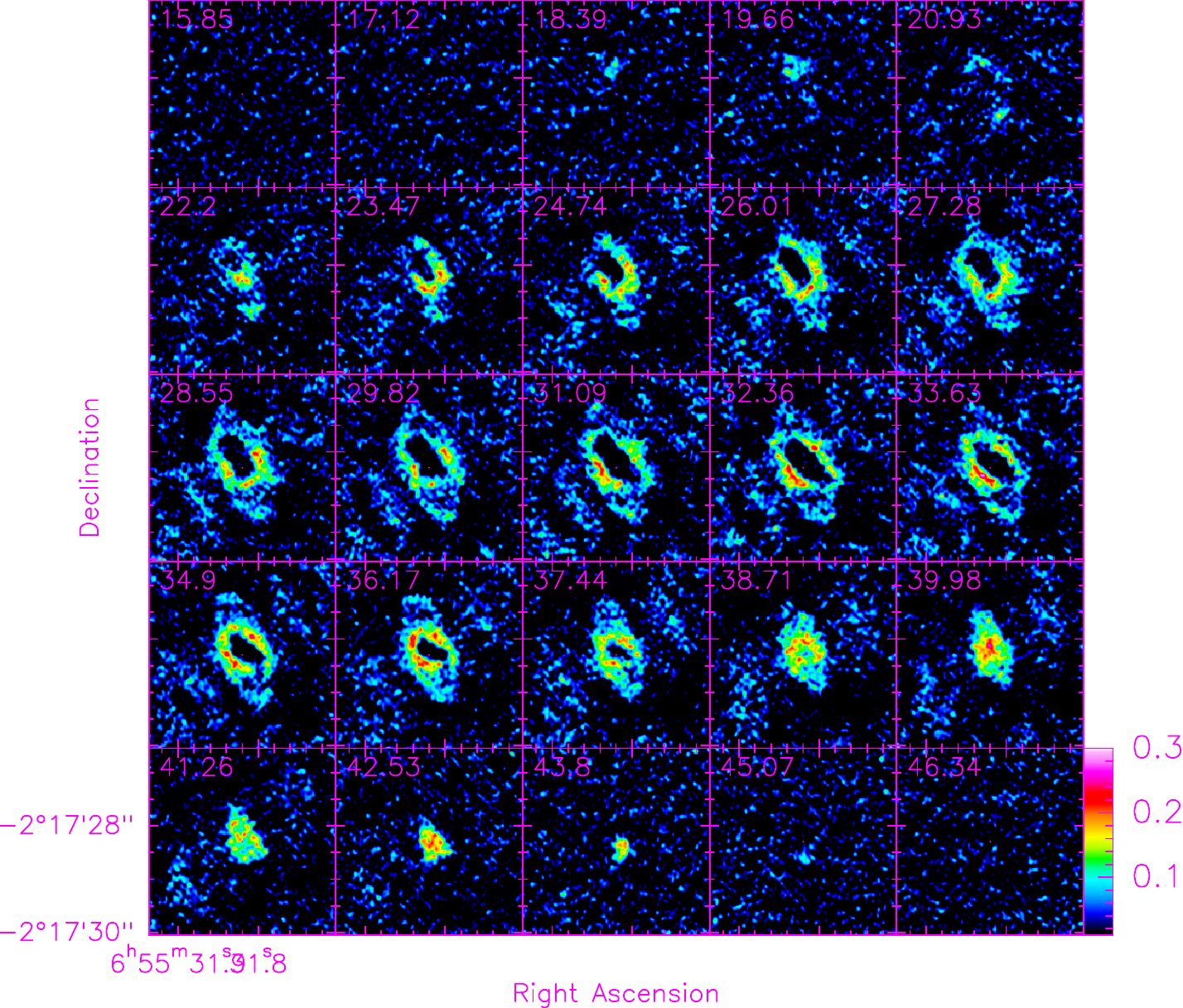}
\caption{Channel maps of the \codos\,\jsix~emission (TM1) towards IRAS\,06530 showing the central bipolar structure. Each channel is 1.27\,\kms~wide, and beam is $0\farcs10\times0\farcs094$, $PA=33.6$\arcdeg. Scale bar shows intensity in units of Jy\,beam$^{-1}$.}
\label{i06530_co65_tm1_ch}
\end{figure}

In Band 9, the CBN is seen in \codos\,\jsix~emission with a spatio-kinematic morphology roughly similar to that seen in \codos\,\jtres, but with lower S/N (Fig.\,\ref{i06530_co65_tm1_ch}) -- the major difference is that the emission from the inner lobe-pair appears very well-separated from that of the much fainter outer lobe-pair, clearly showing that the axis of the former is distinctly different from the latter. We determine a position angle of the inner lobes' symmetry axis of ${\rm PA}=38\pm4$\arcdeg~from the image at the systemic velocity.

\begin{figure}[htbp]

\includegraphics[width=16cm]{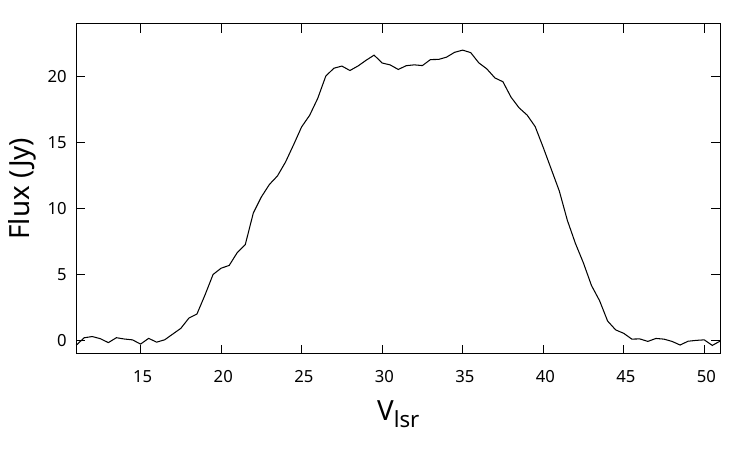}
\caption{Spatially-integrated \codos\,\jsix~spectrum of IRAS\,06530, extracted from the TM1+TM2 datacube, using the CBN aperture.}
\label{co65spec}
\end{figure}

A spectrum of the spatially-integrated \codos\,\jsix~emission from the CBN (Fig.\,\ref{co65spec}) shows a profile that is roughly flat in the central 10\,\kms~velocity range, and sloping wings outside this range. Weak \coeksaat\,\jsix~emission is also detected from the CBN, and appears to be mostly confined to the torus region (Fig.\,\ref{i06530_13co32_h13cn_cs76_mom0}d).

In order to isolate the weak FAR emission, we average the TM1+TM2 \codos\,\jtres~image intensity at the systemic velocity over two azimuthal wedges covering the position angle ranges $-107\arcdeg \le PA \le -35\arcdeg$ and $73\arcdeg \le PA \le 145\arcdeg$ (green curve, Fig.\,\ref{i06530_rcut_co32_tm1}c). These wedges cover the range of position angles where the CBN has the smallest radial extent, allowing us to trace the emission as close as possible to the CBN.  As we move inwards from the ring (i.e., towards small radii), the resulting radial intensity cut shows a slow rise starting at about $2\farcs07$ (i.e., where the intensity exceeds $5\sigma$, with $\sigma$ being the noise in the $\sim2\farcs5-3\farcs5$ region, where no emission can be seen) and then a much faster one starting at about $1\farcs45$; the intensity peaks at about $1\farcs15$, followed by a plateau until $\sim0\farcs98$, where it rises sharply again, due to emission from the bright CBN. A 3-gaussian fit to the emission in the radial interval $0 <$ r($\arcsec$) $< 3.5$ shows that the weakest component arises in the FAR, located at a radius of $1\farcs14$, with a FWHM of $0\farcs48$ (magenta curve, Fig.\,\ref{i06530_rcut_co32_tm1}), and peak intensity of $7.2$\,K. A similar intensity cut for the \codos\,\jsix~reveals weak emission ($\sim1.4$\,K) from the FAR.

\subsection{Continuum Emission}\label{contdata}

The TM1+TM2 band 7 and 9 continuum images show weak, noisy, emission that roughly coincides with the bright trapezoid-shaped periphery of the central cavity seen in \cotres\,\jtres~(Fig.\,\ref{i06530_bipolar}c) with the brightest parts being located roughly in the waist of the CBN, however the waist structure is not as well defined as, e.g., in \hctresn~emission (Fig.\,\ref{i06530_bipolar}b). The total continuum fluxs in band 7 and 9 are, respectively,  $5.5$\,mJy and $54$\,mJy, derived from integrating the emission in the TM1+TM2 image (using an elliptical aperture with semi -major and -minor axes of $1\farcs25$ and $0\farcs65$, respectively, and major axis oriented along ${\rm PA}=22\arcdeg$: hereafter CBN aperture) that adequately covers the emitting region\footnote{The total flux densities derived from an aperture that is twice as large are not significantly different}. We also extracted waist fluxes using a \enquote{waist-aperture} -- a rectangular aperture of width and height equal to $1\farcs3$ and $0\farcs5$, centered on the central star and oriented along the CBN waist (i.e., along ${\rm PA}=22\arcdeg$); these fluxes are $2.3$\,mJy ($22$\,mJy) for band 7 (band 9). Conservative systematic uncertainties in the continuum fluxes are about $\pm10$\%. No central source of compact emission can be seen at the location of the central star in the TM1+TM2 images, or in the highest-angular resolution ($\sim0\farcs1$) TM1 image.

No ring structure can be seen directly in the continuum images in either band 7 or 9. Since averaging over all azimuthal angles can result in a significant increase in the sensitivity to low-level emission, we have made such a radial cut of the band 7 continuum intensity (centered at the central star), and calculated the cumulative flux, F$_{cum}$ as a function of radius (Fig.\,\ref{dust_int_cum}). We find that F$_{cum}$ shows a steep rise from the center to a radius of $r\sim1\arcsec$, followed by a relatively slow increase up to $r\sim3\farcs6$, and then a sharp rise again up to $r\sim6\arcsec$, and then flattens out. The first sharp increase is due to the dust emission from the CBN (and its immediate surroundings), and the second sharp increase is likely associated with dust in the ring. Emission in the region in between these two regions, i.e., $r\sim1\arcsec-3\farcs6$, likely includes a potential contribution from dust associated with the gas in the circumstellar region between the CBN and the ring seen in the form of faint azimuthally-asymmetric structures (see \S\,\ref{struct} and Fig.\,\ref{i06530_co32_azimuthal}).

\subsection{Flux Losses \& Single-Dish Data} We have also obtained observations of the \codos\,\jtres~line using the SMT\,10m telescope in order to estimate flux losses. The data were obtained using both polarisations of the 0.8\,mm receiver, and backends with 1\,MHz and 0.25\,MHz channels. The observed line has a flux of $\sim14$\,Jy at the line-center

\footnote{The beam-efficiency is taken to be 45.5\%, from measurements for April 2023, as listed by the SMT/ARO observatory, https://aro.as.arizona.edu/?q=beam-efficiencies. The main-beam temperature to flux conversion ratio of 46.5 Jy/K}, in good agreement with that measured from the TM1+TM2 ALMA data (Fig.\,\ref{co32specsmt}), implying that the flux losses in the latter are relatively small.

\section{Analysis}\label{analysis}
In this section, we estimate the luminosity of IRAS\,06530's central star, and the temperatures and masses of dust and gas in its various structural components, together with the expansion ages and mass-loss rates of these components. We derive abundance ratios of specific isotopologues of CO and HCN. We set constraints on a potential binary companion to IRAS\,06530's central star. The derived luminosity and masses scale as $D^2$, whereas the expansion ages and mass-loss rates scale as $D$.

\subsection{Bolometric Flux and Luminosity}\label{fbol-lum}
We first determine IRAS\,06530's luminosity (required for other aspects of our analysis of this object). We have used the Vizier archive to extract the photometry for IRAS\,06530 (Table\,\ref{tbl-sed}) to construct its spectral-energy-distribution (SED)\,(Fig.\,\ref{i06530_sed}). Integrating the SED  gives a bolometric flux of $F_{bol}=5.0\times10^{-9}$\,\bfluxu. Correcting the SED for interstellar extinction, using a reddening value of $E(B-V)=1.1$ as derived by \cite{Jayasinghe18,Jayasinghe20}, we get $F_{bol}=5.37\times10^{-9}$\,\bfluxu.

We compare our value of $F_{bol}$ uncorrected for bipolarity, with those given by, or inferred from, the luminosities and adopted distances by other authors. Our value is in good agreement with that derived by \cite{Vickers15}, who model the SED with multiple blackbody curves, and integrate under the overall fit to determine $F_{bol}=5.17\times10^{-9}$\,\bfluxu, but lower than in three other studies. First, \cite{Kamath22} find $L_*=4687$\,\ls~by fitting the optical to near-IR SED with a reddened atmospheric model and distance of $3.8$\,kpc which implies $F_{bol}=6.82\times10^{-9}$\,\bfluxu~-- these authors derived the total interstellar and circumstellar reddening, $E(B-V)=1.85 (+0.02,-0.22)$ from least-squares fitting of the short-wavelength part of the double-humped SED\footnote{this procedure neglects the contribution of scattered light to the optical--near-IR photometry}. Second, \cite{Tosi23} estimate $L_*=6900$\,\ls~using a DUSTY model fit to the SED; their value of the assumed distance, not explicitly provided, was 4.577 kpc (Tosi, S., priv. comm.), which implies that $F_{bol}=1.05\times10^{-8}$\,\bfluxu, significantly higher than our value. The reason for this discrepancy is unknown. Third, \cite{mishra16} derive $L_*=8317$\,\ls~for a distance of 4.7\,kpc, implying $L_*=5374$\,\ls~at our adopted distance of 3.5\,kpc -- this discrepancy appears to be due to the flux values in the SED plotted by \cite{mishra16} being much higher than the observed ones (\S\,\ref{app:sed}).

For bipolar nebulae, the observed bolometric flux can be a factor $1.59$ to $0.63$ times the true bolometric flux, depending on orientation, from pole-on to edge-on, as shown by 2D axisymmetric modeling of the SED of a bipolar PPN \citep{Su98}. Since IRAS\,06530 is seen nearly edge-on, its observed bolometric flux is an underestimate of the true bolometric flux. Applying corrections based on \cite{Su98}, we get $F_{bol}=7.36\times10^{-9}$\,\bfluxu~and $L_*=3285$\,\ls. Taking into account the $F_{bol}$ values from \cite{Kamath22} and \cite{Tosi23}, we assign an uncertainty of $\pm15$\,\% to $F_{bol}$.

\subsection{Dust Emission and Mass}\label{dustmodel}

The continuum emission seen at 0.88 and 0.44\,mm from the CBN arises from thermal dust emission. We can rule out free-free emission from ionized gas since the central star of IRAS\,06530 is not hot enough to produce any substantial ionization. The power-law exponent of the dust emissivity, $\kappa(\lambda)\propto\lambda^{-\beta}$, between 0.44 and 0.89\,mm, derived from the ratio of total continuum fluxes at these wavelengths, assuming a constant dust temperature $T_d$, and that the emission is in the Rayleigh-Jean (R--J) limit and optically-thin (for which $F_\nu = (2 k T_d / \lambda^2) \kappa_\nu / D^2$, \citealt{Hildebrand83}), is 3.2 for the CBN -- implying a dust emissivity power-law index of $\beta=1.2$. In order to estimate $\beta$ (and the total dust mass) more accurately, we model the dust emission source to be a prolate ellipsoidal shell without assuming R--J, with an overdense waist of thickness $0\farcs5$. The outer boundary of the shell is set to be the same as that of the CBN aperture used to extract the flux. The inner radius of the shell is poorly defined since the dust emission is quite noisy. Hence, using the molecular-line emission images as a guide, we allow the inner radius to vary in the range $0\farcs16-0\farcs20$.
The dust density is assumed to vary as $\theta^{-p}$ in the shell; and the overdensity factor of the waist region is taken to be $f_{over}$; we constrain both $p$ and $f_{over}$ from fitting the observed CBN and waist fluxes. We compute the dust temperature at any point within this shell located at an angular distance, $\theta$ from the central star, using:
\begin{equation}
T_d(\theta) = (F_{bol}\,T_{*}^\beta / 4\,\sigma\,\theta^2)^{1/(4+\beta)}
\label{eqn_Td}
\end{equation}
where $T_*$ is the effective temperature of the central star and $\sigma$ is the Stefan-Boltzmann constant\footnote{$\theta$ is in radians}. $T_d(\theta _t)$ is independent of the distance to the source. Taking $T_*=7350\,K$ (appropriate for its spectral type; see \cite{Desmedt16} who find $T_*=7375$\,K$\pm125$\,K), we numerically integrate the resulting emission at 0.44 and 0.89\,mm from each location in the shell, in order to fit the observed continuum fluxes of the CBN and its waist. The maximum $p$ value that produces fits to the observed CBN and waist fluxes is $1.3$, however we rule it out because there is no overdensity in this model, i.e., $f_{over}=1$, whereas the \hctresn~and \cssiete~emission lines (that trace dense molecular gas) clearly shows the presence of an overdense waist region. For lower values of $p=1-0$, we find fits with maximum overdensity factors of $f_{over}=1.2-1.9$. The emissivity index is $\beta=1.29$ ($\sigma=0.04$)\footnote{$\sigma$ is the standard deviation of values of all models that fit the observed fluxes, accounting for the errors in the latter}. The total dust-mass of the CBN is, $M_d(CBN)=(0.38-0.40)\times10^{-3}$\,\ms~($\sigma=0.02\times10^{-3}$\,\ms), assuming an absolute value of the 1\,mm dust emissivity\footnote{the value of $\kappa$ at (sub)millimeter wavelengths is uncertain by at least a factor few, see \cite{setal11vla}} of $\kappa(1mm)=1.5$\,cm$^2$\,g$^{-1}$.

The derived dust mass is inversely proportional to $\kappa(1mm)$. The dust temperature lies in the range $\sim150-255$\,K in the equatorial plane of the shell, and $\sim115-200$\,K along its long axis, with average values of $\sim200$\,K and $\sim180$\,K. The dust mass is relatively insensitive to variations in the inner radius of the shell, since the overall average dust temperature does not vary significantly as a result of such variations. Simply assuming an average dust temperature for the CBN region, 179\,K (derived from our detailed modeling), results in a mass of $M_d = 0.35\times10^{-3}$\,\ms, similar to that derived from the more detailed modeling above. \cite{sun25} measure a 1.3\,mm continuum flux of 1.69\,mJy (with a total uncertainty of 0.48\,mJy) from NOEMA observations of IRAS\,06530 (beam size $0\farcs72\times0\farcs41$)\footnote{presumably using an aperture that cover the full emission region}, and by fitting the cool component of the SED of IRAS\,06530 with dust emission from a cool shell, derive $\beta=1.3$, which agrees well with our value. The predicted 1.3\,mm flux from our modeling, with $\beta=1.29$, is 1.4\,mJy, which is consistent with the NOEMA value within uncertainties. The torus dust mass is $(0.16-0.17)\times10^{-3}$\,\ms~($\sigma=0.9\times10^{-5}$\,\ms), for $p=1-0$.

\begin{figure}[htbp]
\rotatebox{270}{\includegraphics[width=8cm]{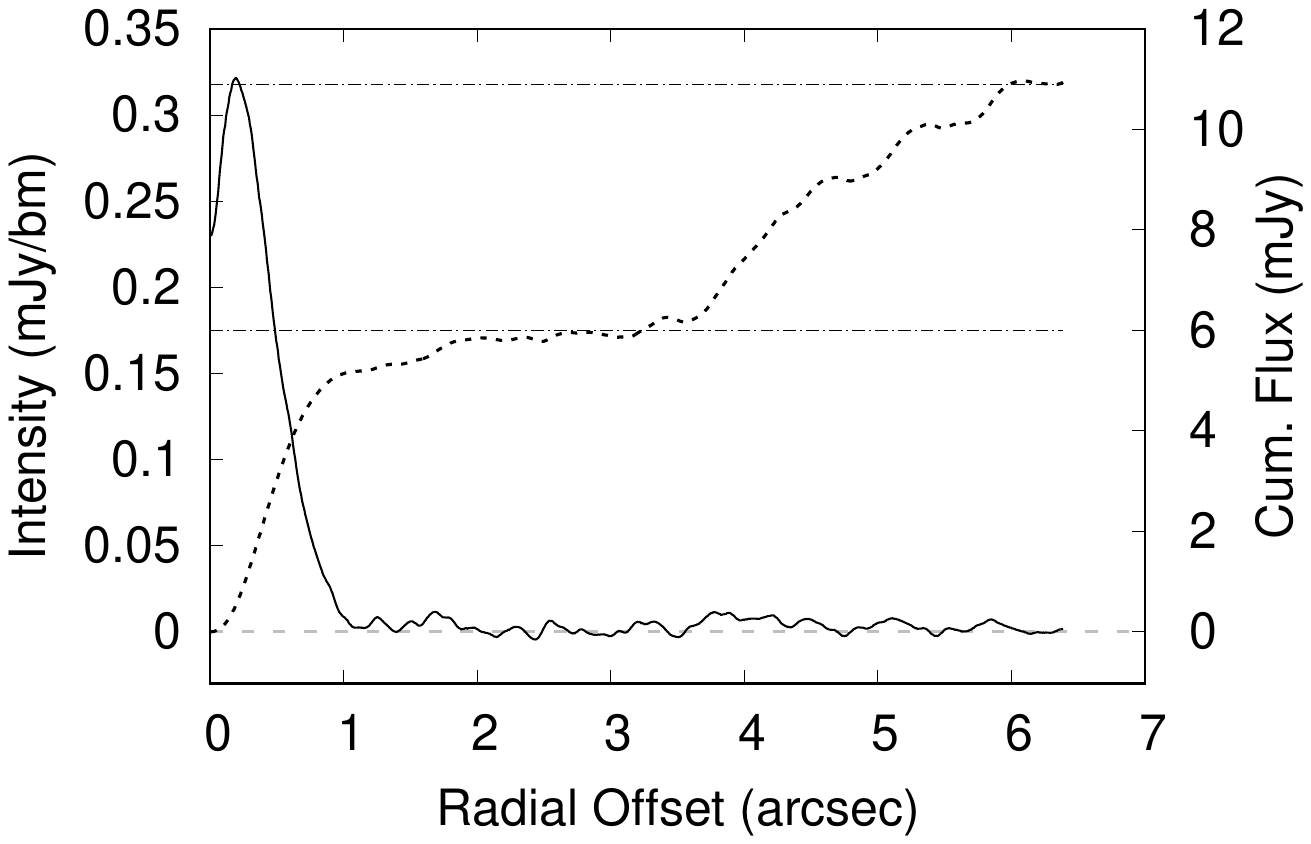}}
\caption{A radial cut of the 0.89\,mm continuum intensity (centered at the central star) extracted from the band\,7 TM1+TM+ACA image by averaging over all azimuthal angles (solid curve), as well as the cumulative flux (dashed curve) as a function of radius. The beamsize is $0\farcs26\times0\farcs22$. Two horizontal dashed lines indicate the cumulative flux level for emission from material interior to the ring ($6$\,mJy), and the total cumulative flux ($10.9$\,mJy).
}
\label{dust_int_cum}
\end{figure}

\begin{figure}[htbp]

\includegraphics[width=16cm]{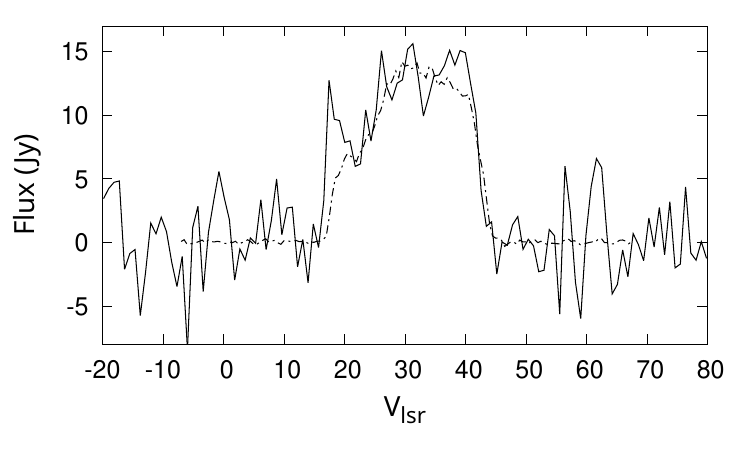}
\caption{Single-dish (SMT) \codos\,\jtres~spectrum of IRAS\,06530 (solid curve), together with a spatially-integrated \codos\,\jtres~spectrum extracted from the TM1+TM2 datacube using a circular aperture of diameter $11\farcs5$ (dash-dot curve).}
\label{co32specsmt}
\end{figure}

We estimate the mass of dust contributing to the ``ring" emission as follows. The maximum flux at the end of the steeply rising 0.89\,mm cumulative flux at $r\gtrsim6\arcsec$ ($10.9$\,mJy: Fig.\,\ref{dust_int_cum}) minus that at the end of its slow increase at $r\gtrsim3\farcs6$ ($6.0$\,mJy) provides a lower limit to the flux emitted by the ring, $4.9$\,mJy. This is a lower limit because it excludes the  contribution (albeit small -- $\lesssim2$\,\% of total) from the ring that lies in projection along the $los$ within an aperture of radius $r\lesssim3\farcs6$. Using equation \ref{eqn_Td}, we find that the dust temperature at radii $r\sim3\farcs6-6\arcsec$ is $(64-78)$\,K, and the corresponding ring dust mass is $M_{d,ring}>1.3\times10^{-3}$\,\ms, assuming the same values of $\beta$ and $\kappa$ as for the CBN.

\subsection{Molecular Masses}\label{tot_mass}

We derive the physical conditions, the total molecular mass (i.e., mass of H$_2$), and the \codos/\cotres~abundance ratio when possible, for the various substructures of IRAS\,06530 using the following general approach. We have generated a large grid of radiative transfer models for \codos~and \cotres~using the RADEX source code with the Large-Velocity-Gradient (LVG) option covering a large range of plausible physical conditions ($T_{kin}$ and $n_{H_2}$) and column densities (divided by the intrinsic line-width, i.e., $N_{CO} / \delta V$ and $N_{{^13}CO} / \delta V$). For each structure we extract the average line intensities, as well as the total line velocity-integrated fluxes, and then find all models that fit these observational constraints, assuming conservative uncertainties for each of these. For the ring and the FAR, an additional ``geometrical" constraint is applied (see \S\,\ref{ringmass}). In addition to our ALMA \jtres~and \jsix~data, we have also used the \codos~and \cotres~\jdos~data from \cite{sun25}. The model grid properties are given in \S\,\ref{app:radex} and Table\,\ref{app:comods}. Note that the upper end of the $T_{kin}$ range is set to 500\,K which is the maximum value for which the RADEX models are valid. Line opacity effects are fully accounted for in our modeling. The derived physical properties are summarised in Table\,\ref{tbl-physprop}.

\subsubsection{The Central Bipolar Nebula (CBN)}\label{cbn_mass}
The details of our modeling strategy that lead to the final values of the physical parameters for the CBN are described in  \S\,\ref{app:modCBN}. We find a kinetic temperature of $\sim57$\,K, a density of \moldens$=7.8\times10^4$, and a \codos/\cotres~abundance ratio of $24$.

The molecular mass is $2.2\times10^{-2}$\,\ms\footnote{all circumstellar masses and mass-loss rates refer to the molecular component, unless noted otherwise}, and the total gas mass, assuming a cosmic abundance of He, is a factor 1.34 higher than the molecular mass, i.e., $M_g = 3\times10^{-2}$\,\ms.

The uncertainties in these quantities derived from our modeling are estimated to be $\lesssim10$\,\%.

The total molecular mass depends on the value of the relative \codos~abundance CO/H$_2=f_{CO}$ (assumed to be $4\times10^{-4}$), which is not well known in PPNe, but is likely to be less than that in the CSEs of the progenitor AGB stars, due to dissociation by shocks resulting from the interaction of fast collimated winds with the ambient CSE material\footnote{this interaction is believed to generate the collimated lobes in these object}. In addition, there may be photodissociation by high-energy radiation (UV and X-rays) resulting from accretion processes that are presumably responsible for producing an accretion disk around a companion, and that power the fast collimated winds. A determination of $f_{CO}$ in an AGB star is difficult, requiring self-consistent modeling that incorporates both radiative transfer and heating+cooling processes. Such modeling gives $f_{CO}=6\times10^{-4}$ for the C-rich star, IRC+10216 \citep{kwan82}, and $f_{CO}=(2-6)\times10^{-4}$ for the C-rich star, U\,Cam \citep{sahai90}. Full association of all the oxygen into CO, assuming a cosmic O abundance, provides an upper limit of, $f_{CO}=10^{-3}$.

Using the dust mass derived in \S\,\ref{dustmodel}, we find that the resulting gas-to-dust ratio is $g2d = M_g/M_d\sim77$, lower than representative values for the gas-to-dust ratio in C-rich AGB stars, $g2d=155$ \citep{knapp85}\footnote{as corrected by \cite{Jura86}} and $g2d=220$ \citep{Jura86}, although even lower values have been derived for individual objects, e.g., in the C-rich PPN, CRL\,2688, \cite{sopka85} find $g2d = 31$. Relatively low gas-to-dust ratios require a relatively high intrinsic C/O ratio, e.g., if $g2d\sim77$, then C/O$\sim2.4-2.7$, assuming solar metallicity, and that the dust is composed of (i) all excess C that is not incorporated into CO (with $f_{CO}=(0.4-1)\times10^{-3}$) and (ii) all refractory elements. For a more plausible value of C/O (e.g., $2$), the gas-to-dust ratio is $g2d\sim100-135$. It is therefore likely that the value of the absolute dust emissivity is somewhat higher, or the relative CO abundance is somewhat lower, than our assumed values (easily accomodated within the uncertainties of our knowledge of these parameters), or some combination of these.

\subsubsection{The Torus in the CBN}\label{torus_mass}
We define the torus region for the molecular gas as a rectangular region with sides of $1\farcs0$ and $0\farcs5$, with the long side oriented along position angle, ${\rm PA}=22\arcdeg$ (hereafter ``torus aperture"), that encapsulates the former as seen in the \hctresn~image at the systemic velocity.
We estimate the molecular mass of the torus in two different ways. First, we assume that the relative abundance of \cotres, $^{13}$CO/H$_2$, in the torus, is the same as for the CBN overall, and since the \cotres\,\jtres~emission is optically-thin, we equate the ratio of the \cotres\,\jtres~integrated intensity extracted from the torus aperture to that from the CBN, $0.33$, to the fractional mass of the torus relative to the CBN mass -- the resulting molecular mass of the torus is $0.73\times10^{-2}$\,\ms. Second, we assume that the gas-to-dust ratio in the torus is the same as for the CBN overall, and therefore the ratios of the torus gas and dust masses, relative to that of the CBN, are equal. The torus dust mass as a fraction of the total CBN dust mass, $0.235-0.30$ for $p=0-1$ (derived from our dust modeling)  results in a molecular mass of the torus of $(0.52-0.66)\times10^{-2}$\,\ms.

We therefore constrain the torus molecular mass to lie in the mass range of $(0.52-0.73)\times10^{-2}$\,\ms, and use the \codos~and \cotres~data to derive the physical parameters of the torus. The details of our modeling strategy that lead to their final values are decribed in \S\,\ref{app:modtorus}. We find that the kinetic temperature and density lie in the range $\sim64-85$\,K and \moldens$=(9.6-32)\times10^4$, respectively. The \codos/\cotres~abundance ratio is $15-21$. Using the average intensities of the \coeksaat\,\jsix~and \cotres\,\jtres~lines in the torus, we find that the \coeksaat/\cotres~abundance ratio lies in the range $0.83-4.9$ (\S\,\ref{app:modtorus}).

We determine an \hcnekpanch/\hctresnmol~abundance ratio in the torus, $0.057$ (with an error of $\sim\pm10$\,\%), using the ratio of the mean \jchar~line emission intensities from these species, $0.056$ (derived by averaging, as above, over a $\sim21$\,\kms~range around the systemic velocity). We find that the $^{14}$N/$^{15}$N ratio in IRAS\,06530's torus lies in the range $254-377$, by combining the \hcnekpanch/\hctresnmol~abundance ratio with the $^{12}$CO/$^{13}$CO ratio in the torus ($15-21$) (see \S\,\ref{app:modtorus}). Similarly, combining the \coeksaat/\cotres~abundance ratio ($0.39-1.0$) with the $^{12}$CO/$^{13}$CO ratio, we find that the \oekchey/\oeksaat~ratio in IRAS\,06530's torus is $15-54$.

\subsubsection{The Ring}\label{ringmass}

We derive the mass of the ring in each of the two components, by fitting the peak \codos\,\jtres~intensities derived from the the 2-gaussian fits (Fig.\,\ref{i06530_rcut_co32_tm1}), assuming a 10\% uncertainty, and the ($3\sigma$) upper limit on the \codos\,\jsix~intensity ($0.55$\,K). We use a similar methodology as for the CBN, but with an additional ``geometrical" constraint, namely, for a spherically-expanding medium, the ratio $R_{N/n} = (N_{CO} / \delta V) / (n_{H_2}\,f_{CO}) = r_{pk} / V_{exp}$, using Eqn.\,1 from \cite{morris1975} for the length of the emitting column. We assume $f_{CO}=4\times10^{-4}$, as for the CBN. The details of our modeling strategy that lead to the final values of the ring's physical parameters are described in \S\,\ref{app:modring}.

The mass of R-Shrp and R-Plat within $1\times$FWHM is $(0.95-2.2)\times10^{-2}$\,\ms and $(2.4-4.4)\times10^{-2}$\,\ms, respectively. The resulting total ring mass is $0.033-0.066$\,\ms~(with R-Plat containing $\sim 70$\,\% of the mass). The total ring mass is significantly larger than the maximum mass ($0.0075$\,\ms) and median mass ($0.0022$\,\ms) found for the detached shells of C-stars by \cite{schoier2005}.

The gas-to-dust ratio in the ring, using the dust mass derived in \S\ref{dustmodel}, is $>(25-51)$, significantly lower than that in the CBN and in C-rich stars (as discused in \S\,\ref{cbn_mass}). Again, as for the CBN, one possible resolution to this discrepancy is that the value of dust emissivity is higher, and/or the relative CO abundance is lower, than our adopted values. However, there is an additional possibility -- the dust ring contains a contribution of swept-up material from the region surrounding it; we discuss this in more detail in \S\,\ref{ages}.

\subsubsection{The FAR}\label{farmass}
We use the \codos\,\jtres~and \jsix~intensities of the FAR $\pm$\,10\% (30\%), and  the same strategy for modeling as for the ring (\S\,\ref{app:modfar}). We find that the mass of the FAR is $(0.12-0.25)$\,\ms -- it thus dominates all other structural components by mass.

\subsection{Filamentary Arcs: Evidence for a Binary Companion?}\label{binary_arcs}

The filamentary arc structures seen in the FAR (e.g., Fig.\ref{i06530_bipolar}a) are reminiscent of the Archimedian spiral structures seen in many AGB CSEs, both in molecular-line emission (e.g., \citealt{Decin20,Claussen11}) and dust emission (e.g., \citealt{Mauron06}), and are widely believed to result from the presence of a close binary companion (e.g., see review by \cite{Decin21}), and supported by numerical simulations (e.g., \citealt{Kim2017,Kim2019,Edgar08}).

Theoretical simulations of such interactions show that the density profiles close to arc features are expected to be quite steep (see Figs.\, 2, 4, and 9 by \cite{Kim2012}). The very steep observed rise in the \codos\,\jtres~emission intensity in the FAR region towards decreasing radii, roughly $r^{-4.1}$ ($r^{-5.3}$) from the half-intensity (one-third intensity) location to the peak may be indicative of such a rise, or may be due to an increase in the mass-loss rate, or both.

\begin{figure}[htbp]
\includegraphics[width=16cm]{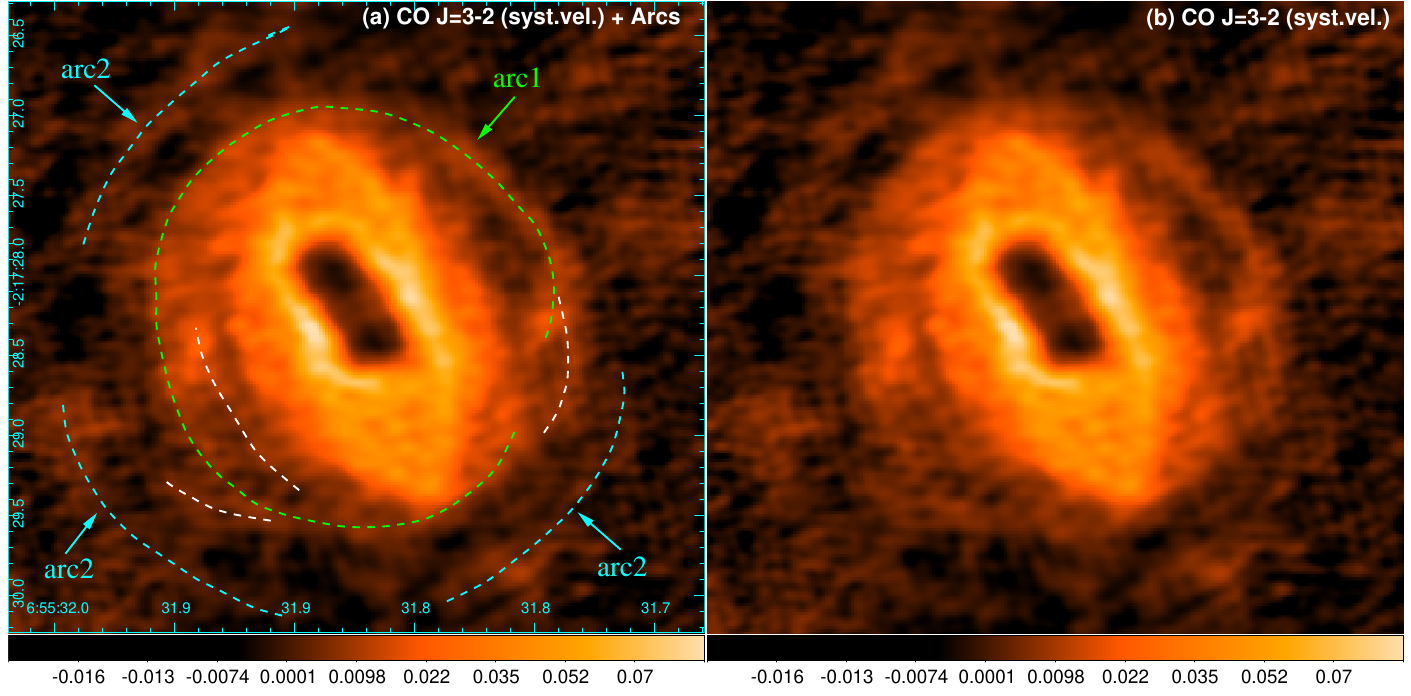}
\caption{(a) \codos\jtres~TM1 image of the central region of IRAS\,06530 at the systemic velocity, showing filamentary arc structures (marked with dashed green, cyan and white arcs) surrounding the central bipolar nebula, (b) same as in panel $a$, but with dashed arcs removed. Scale bars in panels show intensity in units of Jy\,beam$^{-1}$.}
\label{co32_far_arcs}
\end{figure}

This filamentary structure is visible in a large velocity range around the systemic velocity as expected for a 3D spiral structure, and supported by models of binary-induced spiral-shell patterns by \cite{Kim2019}. These models show that the spiral structure does not change near the systemic velocity within offsets that are $\lesssim0.1$ times the expansion velocity (e.g., see model channel maps in Figs.\,7--10 of \cite{Kim2019}). We have therefore averaged the \codos\jtres~emission over the central 2.1\,\kms~velocity range to characterize the filamentary structure. The resulting image shows two prominent arc structures: an outer one (arc\,1) that is relatively faint, appearing as $\sim60\arcdeg$ long threads in the north-east (most prominent), south-east, and south-west quadrants (Fig.\,\ref{co32_far_arcs}), with an average radius of $\sim1\farcs8$ (cyan dashed curves in panel $a$), and an inner brighter one that has an average radius of $\sim1\farcs3$ and is seen over most position angles (arc\,2) (green dashed curve); smaller arcs are also seen (white dashed curves).

The average binary-induced spiral pattern spacing is related to the orbital period of the binary, P. We derive the expansion time-scale between arc\,1 and arc\,2 that are spaced apart by $\Delta r(arc)=0\farcs5\pm0.08$,

$\Delta t_{arc} = \Delta r(arc)\times D/ V_{arc}\,(\kms) = (1220-2900)$\,yr,
where $V_{arc}=9\,\kms$ is the maximum expansion velocity of the arcs derived from P-V plots (e.g., Fig.\,\ref{i06530_co32_tm1_pv}), taking into account the allowed range in $D$ (see \S\,\ref{intro}). Setting P equal to $t_{arc}$, we find a binary separation of $A = (P^2 \, M)^{1/3}$, where $M = M_p + M_c$ is the sum of the primary and companion masses.
Taking $M_p = (1.6-3.4)$\,\ms~(see \S\,\ref{discuss}) and a nominal value of $M_c=1$\,\ms, we get $A = (155-340$)\,au; for $M_c\sim(0.1-2$)\,\ms,  $A\sim(135-355$)\,au (i.e., the separation is relatively insensitive to the value of $M$).

\subsection{Expansion Ages \& Mass-Loss Rates}\label{ages}

The expansion age of the torus is $840$\,yr, obtained by dividing its radial extent, $0\farcs42$ or $2.4\times10^{16}$\,cm\footnote{estimated from the \cotres\jtres~TM2 data averaged over a velocity range of $\Delta V\sim0.89$\,\kms~that spans the central 3 channels around the systemic velocity}, by the torus's expansion velocity $V_{e(wst)}\sim9$\,\kms. The age of the CBN is likely comparable to (but somewhat less than) that of the torus, but it is difficult to make an accurate estimate because we don't know the exact inclination of the symmetry axis, although it appears to be very close to the plane of the sky because spectra extracted from small (rectangular-shaped) regions within the central cavity that are centered on the star, and offset from it along the symmetry axis of the cavity to the north-west and south-east do not reveal any systematic velocity gradient along the axis. Using the expansion velocity of $16.5$\,\kms~derived for a representative location within the lobe walls (\S\,\ref{cbn_desc}), we derive an expansion age there of $580$\,yr, which we assume is representative of the CBN as a whole. The expansion ages of the torus and CBN derived here are upper limits since in both cases, it is likely that their formation involves a faster outflow expanding into a more slowly moving ambient circumstellar medium. We assume that the beginning of the post-AGB phase is marked by the formation of the torus and the CBN, and take the average of their ages as a rough estimate of when the post-AGB phase began, i.e.,  $t_{age(pagb)}\lesssim700$\,yr. The mass-loss rate in the (i) CBN, excluding the torus region, is $>(2.5-2.9)\times10^{-5}$\,\my, and (ii) torus, is $>(6.2-8.7)\times10^{-6}$\,\my.

The expansion ages of the ring's components are obtained by dividing their radii by its expansion velocity ($12.4$\,\kms), i.e., age(R-Shrp) = $6260$\,yr, age(R-Plat) = $6695$\,yr. The mass-loss rates characterising the ring components are obtained by dividing the mass of each component by the corresponding mass-loss duration -- derived by dividing each component's width (FWHM) by the ring expansion velocity. We thus find, for R-Shrp, a duration of $\Delta$t=435\,yr and mass-loss rate of $(2.3-5.1)\times10^{-5}$\,\my; for R-Plat, the duration is $\Delta$t=1155\,yr and the mass-loss rate is $(2.1-3.8)\times10^{-5}$\,\my. However, these estimates are relatively uncertain, when we consider that the inner ring has a relatively narrow width ($0\farcs3$ or $1130$\,au) and comparable to the width resulting from intrinsic expansion over its age, $\sim1300$\,au, due to its intrinsic random velocity dispersion, which is typically $\sim1$\,\kms~in AGB winds. Ruling out the unphysical possibility that the mass ejection that made this ring was virtually instantaneous, its narrow width is likely a result of interaction with a fast tenuous wind that operated after the ring was ejected and that would sweep-up and compress material at the ring's inner boundary. Although the width of the outer ring, $\sim3020$\,au, is large enough that it does not require compression, it is possible that it is being compressed at its outer boundary by interaction with ambient circumstellar material resulting from mass-loss at a smaller expansion velocity prior to the ring ejection, or with the ISM. The relatively low gas-to-dust ratio found for the ring supports such an interaction, because the swept-up material would likely have a low relative CO abundance due to photodissociation (by interstellar UV radiation) if it was due to circumstellar material, or if it was from the ISM (CO/H$_2\sim10^{-4}$ at/near the Galactic plane).

Following the ring-ejection episode, the mass-loss rate decreased and remained very low for about $2145$\,yr, then started increasing, first gradually, and then more rapidly, reaching a peak at $\sim1570$\,yr prior to the post-AGB phase, before declining again. This phase of intense mass-loss, at an average rate of $(1.2-2.7)\times10^{-4}$\,\my, lasted for $\Delta$t$=945$\,yr, and produced the FAR\footnote{taking the expansion velocity of the FAR to be $9$\,\kms, see \S\,\ref{binary_arcs} above}.

The mass-loss rate has likely varied as a function of azimuthal angle in the region between the FAR and the ring, as indicated by the azimuthal variations of the \codos\,\jtres~intensity in this region (\S\,\ref{struct}). We estimate that this mass-loss rate variation is up to a factor $\sim2.5$ between the azimuthally bright and faint regions, using the ratio of the average intensity in the NNW azimuthally bright region to the noise in the diametrically-opposed region with no detectable emission (see \S\,\ref{struct}), and assuming that the emission is optically-thin\footnote{this is reasonable, considering that the average brightness temperature of this region is $\sim 1$\,K, significantly lower than the typical kinetic temperature in a C-rich AGB circumstellar shell at these radii, $\gtrsim10$\,K, e.g., \cite{Huggins88}}.

The expansion ages and mass-loss rates are summarised in Table\,\ref{tbl-age}, together with the scalar momentum and kinetic energy (in order to ease comparison with other studies of PPNe, e.g., \cite{bujarrabal01}).

\section{Discussion}\label{discuss}
The ALMA observations of IRAS\,06530 presented here, provide several new features, some of which are unprecedented and/or quite puzzling. First and foremost amongst these, is the large ring. Similar rings have been seen in C-rich AGB stars, and proposed to have resulted from a thermal pulse (e.g., \cite{Maercker12,Olofsson12}), but never around a post-AGB star. The ring in IRAS\,06530 provides the first direct evidence of the last thermal-pulse immediately preceding the post-AGB phase. The AGB mass-loss rate in IRAS\,06530, \mdot(AGB), appears to have decreased soon after the thermal pulse, and then increased again just before the pAGB phase, declining just prior to the post-AGB phase (see \S\,\ref{ages}). IRAS\,06530's central star was likely still on the AGB just prior to the formation of the CBN.

\begin{figure}[htbp]
\includegraphics[width=6in]{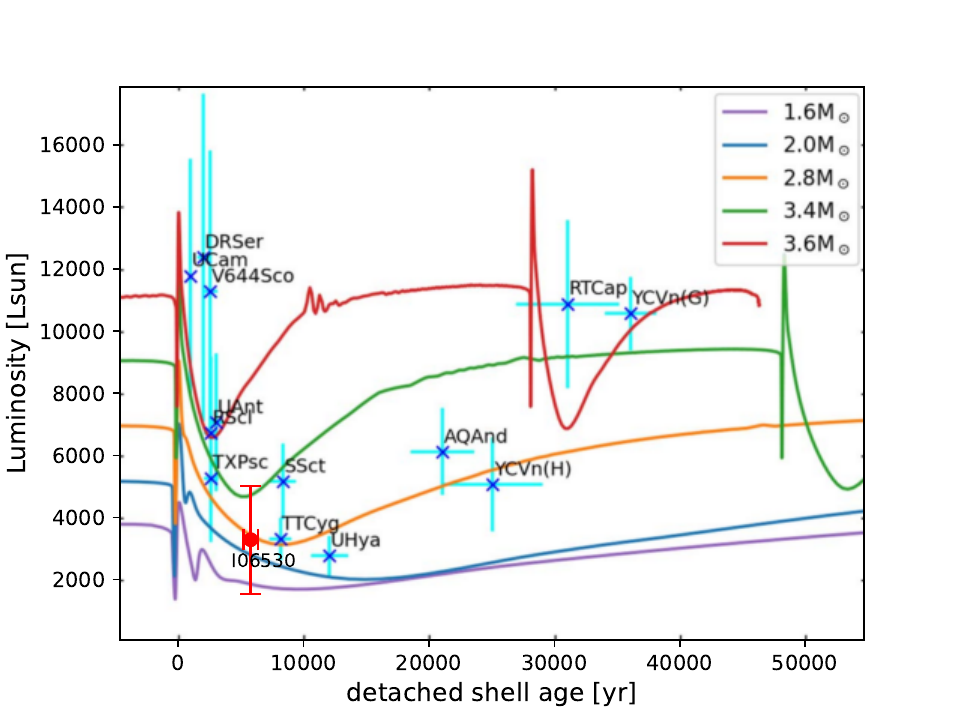}
\vskip 0.1in
\caption{Luminosities of detached-shell carbon stars and IRAS\,06530, versus their shell expansion ages, together with MESA models of the time-evolution of the luminosity during the late AGB thermal-pulsing phase for select progenitor masses in the $(1.6-3.6)$\,\ms~range (adapted from \cite{Kastner2021}.}
\label{tp-shell}
\end{figure}

We compare the luminosity of IRAS\,06530 and the age of its detached shell (while it was still on the AGB), $5780$\,yr\footnote{derived by subtracting IRAS\,06530's post-AGB age from the average age of the ring}, with those of other carbon stars with detached shells, and MESA models of the time-evolution of the luminosity during the late AGB thermal-pulsing phase for progenitor masses in the $(1.6-3.6)$\,\ms~range (Fig.\,\ref{tp-shell}), using Fig. 4 of \cite{Kastner2021}. We find that IRAS\,06530 is located between the tracks for $1.6$\,\ms~and $3.4$\,\ms~progenitors (but closer to the $2.8$\,\ms~track), roughly consistent with the initial mass range, $\sim(2.5-4.0)$\,\ms, found for the majority of C stars known to exhibit detached shells \citep{Kastner2021}, and the average of the range found for Galactic C stars generally, $\sim(1.6-4.5)$\,\ms~\citep{Karakas14, Marigo20}.

If the CBN was formed by a collimated outflow or outflows resulting from binary interaction sculpting the primary's wind from the inside out -- proposed to be the primary mechanism for the formation of aspherical PPNe and PNe \citep{st98,sahai11} -- then these outflows are expected to have had expansion speeds much faster than the primary's wind, i.e., $\gtrsim(10-12)$\,\kms. Most PPNe show direct signatures for such fast outflows in their CO spectra -- the fast moving material, with speeds typically $\gtrsim50$\,\kms, generally resides in the bipolar lobes. So it is rather surprising that we find no evidence for such high velocities in IRAS\,06530's spectra.

We also do not find a compact continuum source at the location of IRAS\,06530's central star in the TM1 continuum image (the $3\sigma$ upper limit is 0.25\,mJy assuming the source is unresolved). Although IRAS\,06530 is not a dpAGB object, it is interesting to compare its lack of a central continuum source with the compact dusty disks (sizes $few\times100$\,au) that are one of the major defining characteristrics of dpAGB objects such as the Red Rectangle and IW\,Car (e.g., \cite{gallardocava21}). The (post-AGB) central star itself, is expected to have a photospheric flux at 0.89\,mm of $3.7\times10^{-3}$\,mJy, and is therefore too faint to be seen in the TM1 image. The $3\sigma$ upper limit for compact disk emission in IRAS\,06530 is, respectively, factors of 88 and 56 lower than the compact disk emission in the Red Rectangle and IW Car, scaled to IRAS\,06530's distance. The absence of a continuum source, together with the presence of the pairs of inner and outer lobes in the CBN (which have likely been produced by collimated outflows that require an accretion-disk engine), suggests that IRAS\,06530 may have harbored a short-lived disk-like structure in the past, which then dissipated after powering a pair of fast collimated outflows for a relatively short period. The operation of the fast outflows for a rather short period would be consistent with the current (observed) lack of high-velocity material in IRAS\,06530.

The presence of the FAR indicates the presence of at least one binary companion (\S\,\ref{binary_arcs}), albeit one with a relatively large separation and therefore somewhat unlikely to be involved in a strong binary interaction with the primary star, unless the orbit is eccentric, allowing the companion to come significant closer to the primary. Various binary iteraction scenarios involving accretion with either MS or WD companions that can generate collimated outflows have been dicussed by \citep{bl14}. Assuming that the aspherical morphology of the CBN is a result of such a binary interaction, we deduce the nature of the binary interaction in IRAS\,06530, using Eqn. 6 of \cite{bl14,bl19} by estimating the accretion rate onto a putative companion, as follows. We assume that the torus and CBN were created roughly at the same time as a collimated jet-like outflow was launched within an ambient CSE -- \cite{Huggins07} finds that jets and torii in PPNe develop almost simultaneously, with jets typically appearing slightly later than tori, with a lag time of a few hundred years. We therefore assume that the relevant acceleration time-scale is the same as the expansion time-scale of the CBN.

We derive the total scalar momentum ($P_{sc}$), kinetic energy ($K.E.$), and the mass-accretion rate of (i) the CBN  (excluding the torus region), taking the average expansion velocity to be $16.5$\,\kms, and (ii) the torus, taking the average expansion velocity to be $9$\,\kms~(Tables\,\ref{tbl-age}, \ref{tbl-accr}). Adding these together, we find (for $D=3.8$\,kpc), a total mass-accretion rate of \mdot$_{accr}=(2.4-2.8)\times10^{-6}\,(Q/2)\,(R_a/M_a)^{0.5}$\,\my, respectively, where $R_a$ and $M_a$ are the radius and mass of the accretor (in solar units), i.e., the companion, and $Q$ is a numerical factor typically satisfying $1 < Q < 5$ in jet models \citep{bl14}. Table\,\ref{tbl-accr} also lists the values of \mdot$_{accr}$ for $D=2.9$\,kpc and $5$\,kpc. The derived values of $P_{sc}$ and $K.E.$ vary as $D^2$, whereas \mdot$_{accr}$ varies as $D$.

Next, we derive the expected Bondi-Hoyle-Littleton (BHL) accretion rate for IRAS\,06530, \mdot$_{BHL}$ using Eqn.\,7 of \cite{bl14}, assuming nominal values for the primary and secondary mass ($M_P=2.5
$\,\ms, $M_S=1$\,\ms), setting the primary's mass-loss rate to be the FAR mass-loss rate with a wind speed $v_w=9$\,\kms. For IRAS\,06530, with a close companion of mass $1$\,\ms, and e.g., an orbital separation of (say) $a_{or}=10$\,au, the BHL accretion rate is $>10^{-5}$\,\my~or $D=2.9-5$\,kpc, i.e., significantly larger than the accretion rate derived from our data (Table\,\ref{tbl-accr}).

For an orbital separation even as large as $a_{or}=75$\,au, the expected accretion rate, \mdot$_{BHL}=(0.95-2.0)\times10^{-6}$\,\my, is comparable to the maximum value derived from our data, if $Q\sim1$. Hence, BHL accretion appears adequate for powering a collimated post-AGB outflow in IRAS\,06530. And since the orbital separation we derived from the FAR substructure is $\sim(155-340)$\,au, we cannot rule out the possibility that the companion responsible for creating the filamentary structure in IRAS\,06530, if in an eccentric object, is also the companion involved in the BHL accretion.

Note that although we don't know the mass or nature of companion (MS or WD) in IRAS\,06530, our inference about what powers its jet engine is likely robust, since the ratio of the accretion radius (assumed to be the same as the radius of the binary companion) to the mass of the accretor (i.e., the companion), $(R_a/M_a)^{0.5}$, varies relatively slowly with $M_a$ for zero-age MS stars ($\propto M_a^{-0.11}$) and terminal age MS stars ($\propto M_a^{-0.25}$) \citep{bl14,bl19}; for a WD, $(R_a/M_a)^{0.5}$ varies as $M_a^{-2/3}$ \citep{kippenhahn13}.

The derived ages of the CBN ($\lesssim580\,(D/3.8\,{\rm kpc})$\,yr) and the torus ($\lesssim840\,(D/3.8\,{\rm kpc})$\,yr), set an upper limit on the post-AGB age of IRAS\,06530, $t_{age(pagb)}$, assuming these were formed at the end of the AGB phase; taking their average of these as a rough estimate, we find $t_{age(pagb)}\lesssim700\,(D/3.8\,{\rm kpc})$\,yr.

\begin{figure}[htbp]
\includegraphics[width=6in]{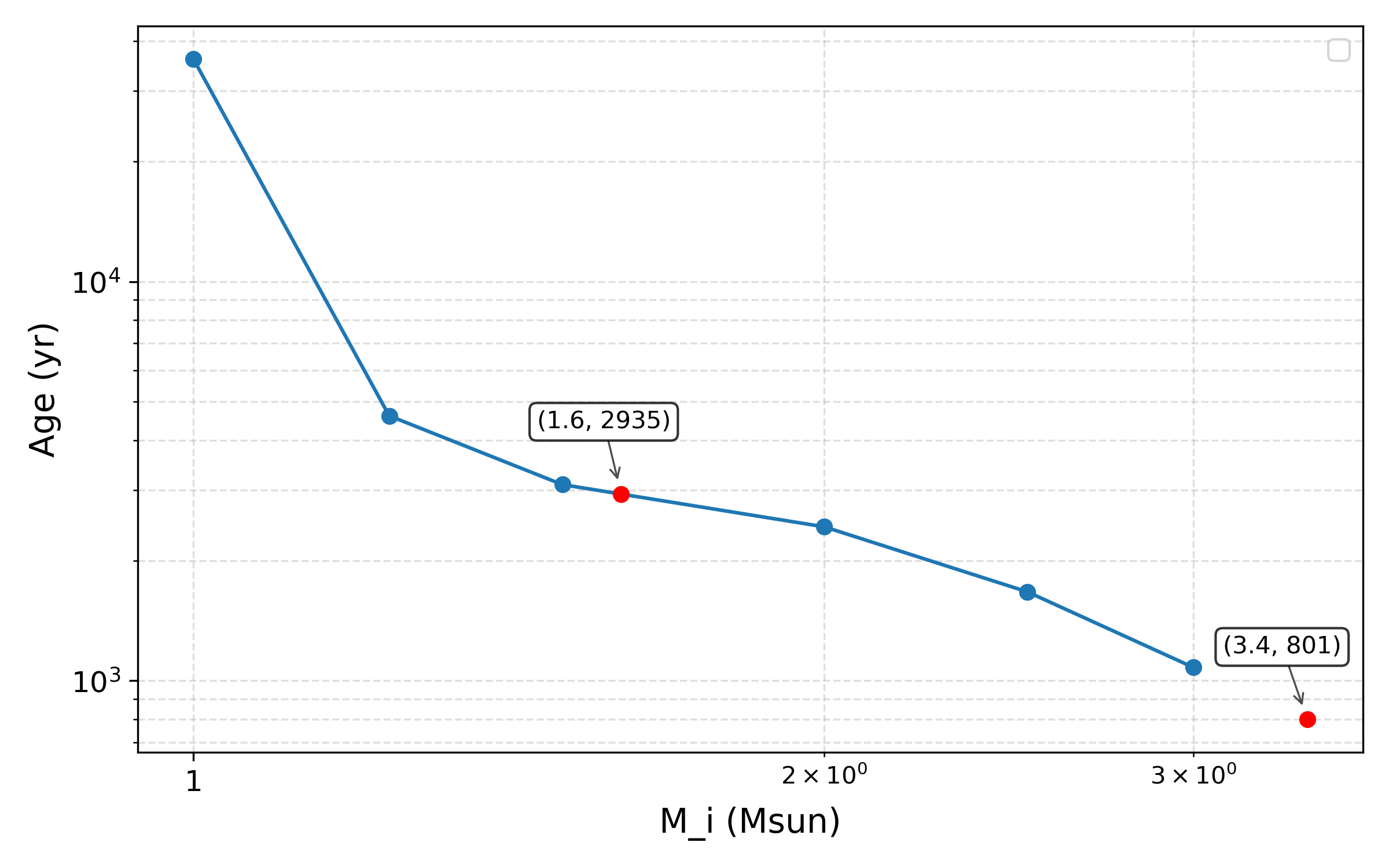}
\vskip 0.1in
\caption{The post-AGB age as a function of the initial stellar mass (M$_i$), using Table\,3 of \cite{bertolami16}. The red symbols mark the interpolated (extrapolated) post-AGB age of IRAS\,06530, assuming its central star had an initial mass of M$_i=1.6$\,\ms~($3.4$\,\ms).}
\label{mi-age}
\end{figure}

The predicted time for a star to evolve from the beginning of the post-AGB phase to an effective temperature of $\sim7100$\,K lies in the range $(2935-800)$\,yr for stars with initial masses of $(1.6-3.4)$\,\ms~and initial metallicity of $Z_0=0.01$\,(appropriate for IRAS\,06530, see \S\,\ref{intro}), based on theoretical models (interpolating/extrapolating values in Table 3 of \cite{bertolami16}) (Fig.\,\ref{mi-age}).

Given IRAS\,06530's effective temperature, $7350$\,K, that is only slightly higher than $7100$\,K, its post-AGB age appears to be significantly smaller than the theoretical predictions (except at or near the top of the range of initial mass values) -- a similar result has been found for a sample of 6 PPNe by \cite{khouri25} who derive post-AGB ages that are a factor of least 2 shorter than the theoretically expected ones. The observed circumstellar gas mass in IRAS\,06530 -- molecular mass of $(0.18-0.34)$\,\ms, implying a total gas mass of $(0.23-0.45)$\,\ms -- represents a significant fraction of its progenitor star's initial mass, as found by \cite{khouri25} for their sample of 6 PPNe.

The value of the $^{12}$C/$^{13}$C~isotope ratio in IRAS\,06530's torus (assumed to be the same as the derived \codos/\cotres~abundance ratio), $18\pm3$, lies below the lower end of the range of values, $25-90$, found in the CSEs of C-rich stars \citep{Milam09}, suggesting that the isotope ratio has decreased in IRAS\,06530 at the very end of the AGB phase. Such a decrease is inconsistent with standard stellar evolution models in which this isotope ratio is expected to steadily increase due to (3rd)dredge-up events and thermal pulsing that adds freshly-formed $^{12}$C into the stellar envelope\footnote{as a result of triple-$\alpha$~nucleosynthesis}. Another C-rich PPN, CRL\,618, similarly shows a low $^{12}$C/$^{13}$C~isotope ratio ($9\pm4$) in the innermost regions of its compact central torus \citep{Lee13}.

IRAS\,06530's torus is significantly enriched in $^{17}$O and $^{13}$C, and modestly enriched in $^{15}$N, when comparing with solar values (Table\,\ref{tbl-isotop}). Similar enrichments have been found for the PN K\,4-47 and J-type carbon stars \citep{schmidt18,abia17,harris87}, the enigmatic object CK\,Vul (believed to have resulted from a stellar merger; \cite{kaminski17}), and in nova grains \citep{lodders05,iliadis18}. Such enhancements cannot be explained by standard models of AGB nucleosynthesis, and non-standard scenarios involving explosive nuclear burning have been proposed \citep{schmidt18,karakas18}.

The mass of molecular ejecta in IRAS\,06530 appears similar to that of CK\,Vul and K\,4-47 ($\sim0.5$\,\ms). However, in contrast to IRAS\,06530, both CK\,Vul (400\,\kms) and K\,4-47 show much higher outflow velocities ($60-80$\,\kms).

\section{Summary}\label{conclude}
We have observed the pre-planetary nebula IRAS\,06530 with ALMA in continuum and molecular-line emission in two wavelength bands (0.44 and 0.89\,mm), with an unprecedented angular resolution of $\sim0\farcs1-0\farcs5$. Our main findings are as follows:
\begin{enumerate}
\item The circumstellar envelope of IRAS\,06530 consists of a remarkable extended circular ring (a geometrically-thin spherical shell in 3-D), a central bipolar nebula (CBN) and torus. The ring (which consists of two close components, with an average radius of $\sim16,950$\,au), provides the first evidence of the last thermal-pulse preceding the post-AGB phase in a carbon star. Assuming the GAIA DR3 parallax distance of 3.8\,kpc to IRAS\,06530, the expansion ages derived for the CBN and torus, assuming these were formed at or just before the beginning of the post-AGB phase, implies that IRAS\,06530 transitioned from an AGB star to a post-AGB star $\lesssim700$\,yr ago.
\item The mass-loss rate of the AGB progenitor of IRAS\,06530 appears to have decreased soon after the last thermal pulse of its progenitor for a period of $\sim2,145$\,yr, and  then increased again just before the post-AGB phase, as evidenced by the presence of filamentary arc structures around the central bipolar source. This filamentary structure is likely part of a 3-D Archimidean spiral shell structure that has been seen in the shells of a small set of C-rich AGB stars and generally attributed to the presence of a binary companion. However, we do not find a compact continuum source at the location of IRAS\,06530's central star, such as that associated with the compact dusty disks (sizes $few\times100$\,au) typically found in disk-prominent post-AGB objects that are known to have binary companions. The $3\sigma$ upper limit for compact disk emission in IRAS\,06530 is, respectively, more than a factor $50$ lower than the compact disk emission in the Red Rectangle and IW Car (scaled to IRAS\,06530's distance).
\item We derive molecular and dust masses for the various structural components of IRAS\,06530 from the line and continuum emission. We find a dust power-law emissivity index ($\kappa(\lambda)\propto\lambda^{-\beta}$), $\beta=1.29$, and a gas-to-dust mass ratio of $77$ in the CBN, significantly lower than the typical values for C-rich AGB stars ($155-220$), but this discrepancy can be resolved by modest adjustments to our assumed values for the absolute dust emissivity ($1.5$\,cm$^2$\,g$^{-1}$ at 1.3\,mm) and/or the CO-to-H$_2$ abundance ratio ($4\times10^{-4}$). The molecular mass budget of IRAS\,06530 is dominated by the filamentary arc region -- $M_{H_2}=(0.12-0.25)$\,\ms~(resulting from a very high average mass-loss rate of $\sim2\times10^{-4}$\,\my); the masses of the ring ($\sim5\times10^{-2}$\,\ms) and CBN ($2.2\times10^{-2}$\,\ms) are much lower.

Faint dust emission from the ring is detected via azimuthal averaging of the radial 0.89\,mm intensity. The derived gas-to-dust ratio in the ring is $>(25-51)$, significantly lower than that in the CBN and in C-rich stars. One possible resolution to this discrepancy is that the value of dust emissivity is higher, and/or the relative CO abundance is lower, than our adopted values (as proposed for the CBN). A lower relative CO abundance would be expected if the ring is expanding into a more slowly moving ambient medium (circustellar or interstellar), and therefore contains a contribution of swept-up material.

\item We derive the isotopic ratios $^{12}$C/$^{13}$C, $^{15}$N/$^{14}$N, and \oekchey/\oeksaat~in the torus. We find that the value of the $^{12}$C/$^{13}$C~isotope ratio ($18\pm3$) is significantly below that found in the CSEs of C-rich stars, suggesting that the isotope ratio has decreased in IRAS\,06530 at the very end of the AGB phase, inconsistent with standard stellar evolution models in which this isotope ratio is expected to steadily increase with time on the AGB. In addition to $^{13}$C, the torus appears to be significantly enriched in \oeksaat, and modestly enriched in $^{15}$N as well (comparing with solar values). This pattern of rare-isotope enrichment, also found in other evolved objects, may result from explosive nuclear burning.
\item We derive a true bolometric flux of $7.36\times10^{-9}$\,\bfluxu~and a luminosity of $3285$\,\ls. A comparison of the luminosity of IRAS 06530 and the age of its detached shell while it was still on the AGB (including the unertainties in each of these), with theoretical  models of the time-evolution of the luminosity during the late AGB thermal-pulsing phase shows that the mass of IRAS 06530's progenitor was $(1.6-3.4)$\,\ms, consistent with the initial mass range found for the majority of C stars known to exhibit detached shells.

\item If the CBN is the result of a collimated post-AGB outflow resulting from binary interaction, then the mass-accretion rate by the companion is \mdot$_{accr}\sim2\times10^{-6}$\,\my. Bondi-Hoyle-Littleton accretion from the progenitor AGB star's wind by the companion appears adequate for powering this post-AGB outflow, even with a relatively large orbital separation ($\sim75$\,au). Thus it is possible that the companion responsible for creating the filamentary structure in IRAS\,06530, if in an eccentric object, is also the companion involved in the BHL accretion.
\end{enumerate}

\section{acknowledgements} 

This paper makes use of the following ALMA data: ADS/JAO.ALMA\#2021.1.00182.S. ALMA is a partnership of ESO (representing its member states), NSF (USA) and NINS (Japan), together with NRC (Canada), MOST and ASIAA (Taiwan), and KASI (Republic of Korea), in cooperation with the Republic of Chile. The Joint ALMA Observatory is operated by ESO, AUI/NRAO and NAOJ. The National Radio Astronomy Observatory is a facility of the National Science Foundation operated under cooperative agreement by Associated Universities, Inc.
 
R.S.’s contribution to the research described here was carried out at the Jet Propulsion Laboratory, California Institute of Technology, under a contract with NASA, and funded in part by NASA via ADAP awards, and multiple HST GO awards from the Space Telescope Science Institute.

C.-F.\,L.  acknowledges support from the National Science and Technology Council of Taiwan (Grant No. 112-2112-M-001-039-MY3).
H.K. was supported by the National Research Foundation of Korea (NRF) grant (No. 2021R1A2C1008928) and the Korea Astronomy and Space Science Institute (KASI) grant (Project No. 2023-1-840-00), both funded by the Korean government (MSIT). G.Q. was supported by the Spanish Ministerio de Ciencia, Innovacioón, y Universidades through grant PID2023-147545NB-I00. J.A., V.B., A.C.-C., were supported by project CRISPNESS, grant PID2023-146056NB-C21, funded by MICIU/AEI/10.13039/501100011033 and by ERDF/EU.
C.S.C. \& M.S.-G. acknowledge financial support from the I+D+i projects PID2023-146056NB-C22 and PID2023-146056NB-C21, supported by MCIN/AEI/10.13039/501100011033.

\scriptsize

\begin{longtable}{cccccccc}

\caption[]{ALMA Molecular Line Observations of IRAS\,06530--0213}\\
\hline \\[-2ex]
   \multicolumn{1}{l}{\textbf{SPW.}} &
   \multicolumn{1}{l}{\textbf{Line}} &
   \multicolumn{1}{l}{\textbf{Freq.}} &
   \multicolumn{1}{l}{\textbf{Range}} &

   \multicolumn{1}{l}{\textbf{$\delta\,\nu$}} &
   \multicolumn{1}{l}{\textbf{Array}} &
   \multicolumn{1}{l}{\textbf{Beam}}  \\

   \multicolumn{1}{l}{\textbf{}} &
   \multicolumn{1}{l}{\textbf{}} &
   \multicolumn{1}{l}{\textbf{GHz}} &
   \multicolumn{1}{l}{\textbf{GHz}} &
   \multicolumn{1}{l}{\textbf{MHz}} &
   \multicolumn{1}{l}{\textbf{}} &
   \multicolumn{1}{l}{\textbf{$\arcsec\times\arcsec$,\,(PA)$\arcdeg$}} &

   \\[-1.8ex]
\endfirsthead

\multicolumn{3}{c}{{\tablename} \thetable{} -- Continued} \\[0.5ex]
\caption[]{Log of Molecular Line Observations}\\
\hline \\[-2ex]
   \multicolumn{1}{l}{\textbf{SPW.}} &
   \multicolumn{1}{l}{\textbf{Line}} &
   \multicolumn{1}{l}{\textbf{Freq.}} &
   \multicolumn{1}{l}{\textbf{Range}} &

   \multicolumn{1}{l}{\textbf{$\delta\,\nu$}} &
   \multicolumn{1}{l}{\textbf{Array}} &
   \multicolumn{1}{l}{\textbf{Beam}}  \\

   \multicolumn{1}{l}{\textbf{}} &
   \multicolumn{1}{l}{\textbf{}} &
   \multicolumn{1}{l}{\textbf{GHz}} &
   \multicolumn{1}{l}{\textbf{GHz}} &
   \multicolumn{1}{l}{\textbf{MHz}} &
   \multicolumn{1}{l}{\textbf{}} &
   \multicolumn{1}{l}{\textbf{$\arcsec\times\arcsec$,\,(PA)$\arcdeg$}} &

   \\[-1.8ex]
\endhead
\\
\hline
\multicolumn{7}{c}{\textbf{Band 7}} \\
\hline
25  & \cotres\,\jtres & 330.58797 & 330.2974--331.233 &  0.4883 & 12m (TM1)  & $0.14\times0.094$, -82.7 \\ 
{''}  &{''}&{''}&{''}&{''}& 12m (TM2)  & $0.51\times0.46$, -43.9 \\

16  &{''}&{''}&{''}&{''}& 7m (ACA)  & $5.5\times2.9$, -87.50 \\
31  & \hcnekpanch\,\jchar & 344.20032 & 344.0303--344.4971 &  0.4883 & 12m (TM2)  & $0.50\times0.44$, -39.3 \\

22  &{''}&{''}&{''}&{''}&  7m (ACA)  & $5.1\times2.8$, -86.9 \\
33  & \cssiete & 342.88285 & 342.6305--343.0974 &  0.4883 & 12m (TM1)  & $0.13\times0.092$, -82.4 \\ 
{''}  &{''}&{''}&{''}&{''}& 12m (TM2)  & $0.5\times0.45$, -42.5 \\
24    &{''}&{''}&{''}&{''}&  7m (ACA)  & $5.2\times2.8$, -86.6 \\

35  & \codos\,\jtres  & 345.79599 & 345.0958--346.0314 &  0.4883 & 12m (TM1)  & $0.13\times0.090$, -82.6 \\ 
{''}  &{''}&{''}&{''}&{''}& 12m (TM2)  & $0.49\times0.44$, -40.2 \\
26    &{''}&{''}&{''}&{''}&  7m (ACA)  & $5.0\times2.8$, -86.4 \\

35  & \hctresn & 345.33976 &{''}&{''}& 12m (TM1) & $0.13\times0.090$, -82.6 \\ 
{''}  &{''}&{''}&{''}&{''}& 12m (TM2)  & $0.49\times0.44$, -40.2 \\
26    &{''}&{''}&{''}&{''}&  7m (ACA)  & $5.0\times2.8$, -86.4 \\

\hline
\multicolumn{7}{c}{\textbf{Band 9}} \\
\hline
41  & \coeksaat\,\jsix & 674.00934 & 673.1834--675.0546 &  0.9767 & 12m (TM2)  & $0.32\times0.28$, 78.3  \\
43  & \codos\,\jsix & 691.47308 & 690.4646--692.337 &  0.9767 & 12m (TM1)  & $0.1\times0.094$, 33.6  \\ 
{''}  &{''}&{''}&{''}&{''}& 12m (TM2)  & $0.32\times0.28$, 85.1  \\

43  & \hctresnhi & 690.55207 & {''} &  0.9767 & 12m (TM1)  & $0.1\times0.094$, 33.6  \\ 

{''}    & {''} & {''} & {''} &  {''}& 12m (TM2)  & $0.32\times0.28$, 85.1  \\ 

\hline

\label{tbl-obslog}
\end{longtable}

\begin{table}
\caption{Line-free frequency ranges used to obtain the continuum maps in TM1 \& TM2 in Band\,7.}
\begin{center}
    \begin{tabular}{lccc}
        \hline
        SPW & & Freq. Ranges\,(MHz) & \\
        \hline
    25&    330333.441--330543.264& 330627.296--331266.472    \\
    27&    332266.753--332731.821& &                \\
    29&    333667.958--333886.139& 333928.134--334133.026     \\
    31&    344066.265--344141.012& 344243.660--344533.742    \\
    33&    342665.989--342855.322& 342915.980--343133.815    \\
    35&    345133.108--345286.753& 345389.460--345748.934& 345851.641--346066.778 \\
        \hline
    \end{tabular}
\end{center}
\label{tbl-cont7}
\end{table}

\begin{table}
\caption{Line-free frequency ranges used to obtain the continuum maps in TM1 \& TM2 in Band\,9.}
\begin{center}
    \begin{tabular}{lccc}
        \hline
        SPW & Freq. Ranges\,(MHz) & & \\
        \hline
    33&    678606.382--676732.342 \\
    35&    686962.492--688836.531 \\
    37&    675149.329--677023.368 \\
    39&    688562.492--690436.531 \\
    41&    673168.126--673867.350&    674031.414--675042.165   \\
    43&    690535.492--691425.148&    691524.758--692409.531   \\
    45&    670841.369--671545.477&    671744.697--672715.409   \\
    47&    692862.492--693483.591&    693614.452--694736.531   \\
        \hline
    \end{tabular}
\end{center}
\label{tbl-cont9}
\end{table}

\begin{table}
\caption{Observed Emission Properties of Structures in IRAS\,06530--0213}
\begin{center}
\begin{tabular}{lccccc}
\hline
Structure\tablenotemark{a} & Line   & <Intensity>\tablenotemark{b}&  $\Delta$V\tablenotemark{c}& <Intensity>\tablenotemark{d} &  Area(Emiss.)\tablenotemark{e} \\
                           &        &(around line-center)        &                             & (velocity-integrated) & \\
Name                       &        &(K)                         &  (\kms)                     & (K\,\kms)             & $10^7$\,au$^2$\\
\hline
CBN                 & \codos\,\jtres &31.9\,($\pm$10\%)          & 1.5                         & ...    &  6.726\\
...                 & \codos\,\jsix  &23.5\,($\pm$10\%)          & 1.5                         & 445.6  &{''}  \\
...                 & \codos\,\jdos  &32.7\,($\pm$15\%)          & 1.5                         & ...   &{''}  \\
...                 & \cotres\,\jtres&2.15\,($\pm$15\%)          & 1.5                         & 54.6  &{''}  \\
...                 & \cotres\,\jdos &0.96\,($\pm$20\%)          & 1.5                         & ...   &{''}  \\
\\
Torus               & \codos\,\jtres &52.0\,($\pm$10\%)          & 1.5                         & ...   &  0.714\\
...                 & \codos\,\jsix  &45.9\,($\pm$10\%)          & 1.5                         & 702.8 &{''}  \\
...                 & \codos\,\jdos  &45.0\,($\pm$15\%)          & 1.5                         & ...   &{''}  \\
...                 & \cotres\,\jtres&4.0\,($\pm$15\%)           & 1.5                         & 92.0  &{''}  \\
...                 & \cotres\,\jdos &2.2\,($\pm$20\%)          & 1.5                          & ...   &{''}  \\
\\
R-Shrp              & \codos\,\jtres &4.0\,($\pm$10\%)          & 1.5                         &...    &...    \\
...                 & \codos\,\jsix  &$<0.55$\tablenotemark{f}  & 1.5                         &...    &...    \\
R-Plat              & \codos\,\jtres &2.2\,($\pm$10\%)          & 1.5                         &...    &...    \\
...                 & \codos\,\jsix  &$<0.55$\tablenotemark{f}  & 1.5                         &...    &...    \\
\\
FAR                 & \codos\,\jtres &7.2\,($\pm$10\%)          & 1.5                         &...    &...    \\
...                 & \codos\,\jsix  &1.4\,($\pm$30\%)          & 1.5                         &...    &...    \\

\hline
\end{tabular}
\end{center}
\tablenotetext{a}{Col.\,1: Name of Structure}
\tablenotetext{b}{Col.\,2: Mean intensity over the emitting region, around line-center}
\tablenotetext{c}{Col.\,3: Velocity-width around line-center used for computing mean intensity in Col.\,2}
\tablenotetext{d}{Col.\,4: Total velocity-integrated mean intensity over the emitting region}
\tablenotetext{e}{Col.\,5: Area of emitting region within aperture used for analysis}
\tablenotetext{f}{$3\,\sigma$ upper limit}
\label{tbl-int-mom0}
\end{table}

\begin{table}

\caption{Physical Properties of Structures in IRAS\,06530--0213: Modeling Constraints \& Results}\label{tbl-physprop}
\resizebox{\textwidth}{!}{

\begin{tabular}{lccccll}
\hline
Structure & $T_{kin}$\tablenotemark{a} & $n_{H_2}$\tablenotemark{b}   & $M_{H_2}$\tablenotemark{e} & \cdos/\ctres\tablenotemark{f} & Constraints & Comments\\
Name      & (K)                        & (${\rm 10^4\,cm}^{-3}$)            & ($10^{-2}\,\ms$)           &             &         \\
\hline
CBN       & $47-57$                    & $7.8-10^3$       & $1.4-3.0$  &    & \codos~lines & max\,(min)\,$n_{H_2}$ at min\,(max)\,$T_{kin}$ \\
CBN(final)& $57$                       & $7.8$            & $2.2$      &$24$& \cotres~lines& \codos\,\jdos~too bright if $n_{H_2}>7.8\times10^4$ \\
\hline
torus     & $60-220$                   & $1.3-10^3$       & $0.33-7.8$ &    & \codos~lines & max\,(min)\,$n_{H_2}$ at min\,(max)\,$T_{kin}$ \\
...       & $60-99$                    & $9.6-14.4$       & $0.33-2.2$ &    & \cotres~lines&  \\
          &                            &                               &    &    & \& $n_{H_2}$(CBN) $< n_{H_2}$(tor) $\le 3\, n_{H_2}$(CBN) & \\
          &                            &                               &    &    & \& $M_{H_2}$(tor) $=(0.52-0.73)\times10^{-2}$\,\ms\tablenotemark{h} &  \\
torus(final)& $64-85$                  & $9.6-14.4$       & $0.52-0.73$& $15-21$ & $M_{H_2}$(tor) $\le 0.33\,M_{H_2}$(CBN) &  \\
...       &                            &                  &            &         & \& $M_{H_2}$(tor) $\ge 0.24\,M_{H_2}$(CBN) &  \\

\hline
FAR       &$>13.7$                     & $8.3-18$       & $12-25$   & ... & \codos~lines, geometrical\tablenotemark{j} & $1\times$FWHM\tablenotemark{k} \\

\hline
R-Shrp    &$>50$\tablenotemark{g}      & $610-1120$     & $0.95-2.2$ & ... & \codos~lines, geometrical\tablenotemark{j} & $1\times$FWHM\tablenotemark{k} \\

R-Plat    &$>50$\tablenotemark{g}      & $500-915$      & $2.4-4.4$ & ... & \codos~lines, geometrical\tablenotemark{j} & $1\times$FWHM\tablenotemark{k} \\

Ring\,(total) &                      &                & $3.3-6.6$ & ... &                                            & $1\times$FWHM \\

\hline
\end{tabular}
}
\tablenotetext{a}{Gas kinetic temperature; values show the allowed range, except in the case of CBN(dust), for which the values reflect the minimum and maximum values present in the emitting region}
\tablenotetext{b}{Volume density; values show the allowed range}
\tablenotetext{c}{Column density / line-width; values show the allowed range}
\tablenotetext{d}{For CBN and Torus: Area of Emitting Region; for R-Shrp and R-Plat, width  (FWHM) of ring component (projected on sky-plane)}
\tablenotetext{e}{Mass of Emitting Region; values show the allowed range, for $D=3.8$\,kpc}
\tablenotetext{f}{Isotope ratio assumed to be equal to the \codos/\cotres~abundance ratio; values show the allowed range}
\tablenotetext{g}{assumed, see text}
\tablenotetext{h}{see \S\,\ref{torus_mass}}
\tablenotetext{j}{see \S\,\ref{ringmass}}
\tablenotetext{k}{gaussian-density profile, integrated over width $1\times$FWHM}

\label{tbl-modfits}
\end{table}

\begin{table}
\caption{Expansion Ages and Mass-Loss Rates of Structures in IRAS\,06530 (for $D=3.8$\,kpc)}
\begin{center}
\begin{tabular}{lccccccc}
\hline
Structure & Radius & V$_{exp}$ & Expansion Age\tablenotemark{a} & Duration\tablenotemark{b} & \mdot\tablenotemark{b} & $P_{sc}$  & $K.E.$\\
Name      &  (AU)  &  \kms   & (yr)                           & (yr)                        & $10^{-4}$\,\my         & $10^{37}$\,\pscalcgs & $10^{43}$ \engyunit \\
\hline
Ring\,(R-Shrp)\tablenotemark{c} & 16,375 & 12.4 & 6260     &  435       & $0.22-0.50$ & $2.3-5.4$  & $1.5-3.4$ \\
Ring\,(R-Plat)\tablenotemark{c} & 17,515 & 12.4 & 6695     & 1160       & $0.20-0.38$ & $5.9-10.8$ & $3.7-6.7$ \\
FAR                             &  4,310 &  9.0 & 2270     &  945       & $1.2-2.7$   & $5.9-11.8$ & $2.7-5.3$ \\
Torus                     &  1,590 &  9.0 & $\lesssim$840 & ...         & ...         & $0.93-1.3$ & $0.42-0.59$ \\
CBN\tablenotemark{d}   &  2,005\tablenotemark{e} & 16.5 & $\lesssim$580 & ...         & ...         & $4.8-5.5$  & $4.0-4.5$ \\
\hline
\end{tabular}
\end{center}
\tablenotetext{a}{At the location of the peak intensity of gaussian fit to local radial intensity \S\,\ref{ages}}
\tablenotetext{b}{Mass-loss duration and rate derived using the FWHM of gaussian fit to local radial intensity \S\,\ref{ages}}
\tablenotetext{c}{R-Shrp and R-Plat represent two components of the ring structure, see \S\,\ref{ring_desc}}
\tablenotetext{d}{CBN with torus region excluded}
\tablenotetext{e}{representative location within the CBN walls, see \S\,\ref{cbn_desc}}
\label{tbl-age}
\end{table}

\begin{table}
\caption{Mass Accretion Rates onto Companion in IRAS\,06530}
\begin{center}
\begin{tabular}{lcccccc}
\hline
Structure & D     &  \mdot         & D     &  \mdot         & D     &  \mdot         \\
Name      & (kpc) & $10^{-6}$\,\my & (kpc) & $10^{-6}$\,\my & (kpc) & $10^{-6}$\,\my \\
\hline
\multicolumn{7}{c}{Mass-accretion rate from structure, onto a $1$\,\ms~companion} \\
\hline
CBN\tablenotemark{a}& 3.8    & $2.1-2.4$   &  5.0   & $2.8-3.2$   &  2.9    & $1.6-1.8$ \\
Torus     & \ditto & $0.28-0.39$ & \ditto & $0.37-0.52$ & \ditto  & $0.21-0.30$ \\
Total\,(CBN+Torus) & \ditto & $2.4-2.8$   & \ditto & $3.2-3.7$   & \ditto  & $1.8-2.1$ \\
\hline
\multicolumn{7}{c}{Bondi-Hoyle-Littleton mass-accretion rate, onto a $1$\,\ms~companion} \\

\hline
BHL(10\,au\tablenotemark{b}) & 3.8\tablenotemark{b} & $15-32$    & 5.0\tablenotemark{d} & $20-44$   & 2.9\tablenotemark{e} & $11-24$ \\
BHL(75\,au\tablenotemark{b}) & \ditto               & $0.95-2.0$ & \ditto               & $1.7-3.6$ & \ditto               & $0.56-1.2$ \\
\hline
\end{tabular}
\end{center}

\tablenotetext{a}{CBN with torus region excluded}
\tablenotetext{b}{Semi-major axis of binary orbit, see \S\,\ref{discuss}}
\tablenotetext{c}{Primary mass $2.5$\,\ms}
\tablenotetext{d}{Primary mass $3.4$\,\ms}
\tablenotetext{e}{Primary mass $1.6$\,\ms}
\label{tbl-accr}
\end{table}

\begin{table}
\caption{Isotope Ratios in IRAS\,06530--0213's Torus}
\begin{center}
\begin{tabular}{lcccc}
\hline
Ratio             & Tracer                                    & Torus& K\,4-47\tablenotemark{a} & Solar\tablenotemark{b} \\
                  & Molecule                                  &                       &         &\\
\hline
$^{12}$C/$^{13}$C & \codos, \cotres                           & $15-21$       & $2.2\pm0.8$  & $89.4\pm0.2$ \\
$^{14}$N/$^{15}$N & \hctresnmol, \hcnekpanch, \codos, \cotres & $254-377$     & $13.6\pm6.5$ & $435\pm57$   \\
$^{16}$O/$^{17}$O & \coeksaat, \codos, \cotres                & $15-54$      & $21.4\pm10.3$ & $2632\pm7$   \\
\hline
\end{tabular}
\end{center}
\tablenotetext{a}{from \cite{schmidt18}}
\tablenotetext{b}{from \cite{asplund09}}
\label{tbl-isotop}
\end{table}

\clearpage

\clearpage
\appendix
\section{Supplemental Information}

\renewcommand{\thefigure}{A\arabic{figure}}

\renewcommand{\thetable}{A\arabic{table}}

\setcounter{figure}{0}
\setcounter{table}{0}

\subsection{Spectral Energy Distribution}\label{app:sed}
\begin{figure}[H]
\includegraphics[width=10cm]{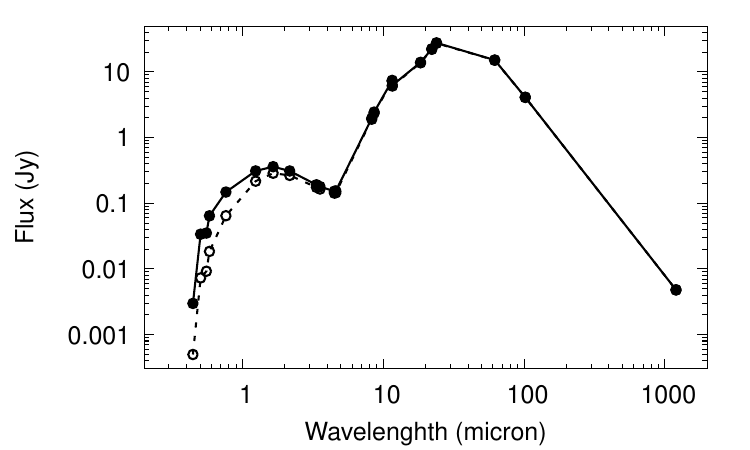}

\caption{The spectral energy distribution of IRAS\,06530. Filled (empty) circles show the photometry from 0.4 to 1200\,\micron~corrected (uncorrected) for ISM-extinction joined by solid (dashed) curves.}
\label{i06530_sed}
\end{figure}

We have extracted the photometry of IRAS\,06530 using the Vizier tool at http://vizier.cds.unistra.fr/vizier/sed, except for the 1.2\,mm flux, which was derived from observations with the IRAM\,30-m MAMBO array by \cite{buemi07}. This photometry, corrected (for ISM extinction, see S\,\ref{fbol-lum}) and uncorrected, is tabulated in Table\,\ref{tbl-sed}, and the resulting SEDs are shown in Fig.\,\ref{i06530_sed}. We note that the SED shown by \cite{mishra16} in their Fig. 2 appears to be incorrect at wavelengths shorter than 1\,\micron, with the flux values being much higher than the observed ones; furthermore the shorter-wavelength peak of their double-humped SED appears to be around 0.7\,\micron, whereas the actual peak is at 1.65\,\micron~(2MASS H-band). The SED plot in \cite{mishra16} is missing the 2MASS fluxes.

\begin{table}[H]

\caption{IRAS\,06530-0213 Photometry}
\begin{tabular}{lccc}
\hline
Wavelength   &  Flux &  Flux(corr.)\tablenotemark{a} & Catalog/Filter\\
($\micron$) &  (Jy) &  (Jy) \\
\hline
0.444 & 0.000498 & 0.00299 & Besssel\,B\\
0.504 & 0.0073 & 0.0338    & Gaia\,Bp\\
0.554 & 0.00923 & 0.0352   & Bessel\,V\\
0.582 & 0.0185 & 0.0645    & Gaia\,G\\
0.762 & 0.0647 & 0.148     & Gaia\,Rp\\
1.24 & 0.217 & 0.311       & 2MASS\,J\\
1.65 & 0.287 & 0.361       & 2MASS\,H \\
2.16 & 0.266 & 0.31        & 2MASS\,Ks \\
3.35 & 0.175 & 0.191       & WISE\,W1 \\
3.55 & 0.165 & 0.179       & IRAC\,Ch.1\\
4.49 & 0.143 & 0.152       & IRAC\,Ch.2 \\
4.6 & 0.145 & 0.154        & WISE\,W2  \\
8.28 & 1.91 & 1.97         & MSX\,A   \\
8.61 & 2.37 & 2.45         & Akari\,S9W\\
11.6 & 7.26 & 7.43        & WISE\,W3   \\ 
11.6 & 6.11 & 6.26        & IRAS\,12 \\
18.4 & 13.8 & 14          & Akari\,L18W\\
22.1 & 22.1 & 22.4        & WISE\,W4  \\
23.9 & 27.4 & 27.7        & IRAS\,25  \\
61.8 & 15.1 & 15.2        & IRAS\,60   \\
102 & 4.1 & 4.11         & IRAS\,100  \\
1200 & 0.0048 & 0.0048 & IRAM\,30m MAMBO \\
\hline
\end{tabular}
\tablenotetext{a}{Corrected for interstellar extinction, see \S\,\ref{fbol-lum}}
\label{tbl-sed}
\end{table}

\subsection{RADEX modeling}\label{app:radex}
As summarised in the main paper, we have used a large library of CO models to derive the physical properties of the various structures in IRAS\,06530. We describe some of the important trends in the variations of the \codos\,\jtres~and \jsix~line intensities as a function of T$_{kin}$ and $n_{H_2}$.

\subsubsection{\codos~\& \cotres~Model Database}\label{app:comods}
\begin{table}[H]
\caption{\codos~\& \cotres~Model Parameter Grid}
\begin{tabular}{lccc}
\hline
Physical          &  Min      & Max     & No. of Grid\\
Parameter         &           &         & points  \\
\hline
T$_{kin}$(K)\tablenotemark{a} &  10     & 500     &   100 \\
$n_{H_2}$ (cm$^{-3}$)  & $5\times10^2$   & $10^7$ &   100 \\
$N(^{12}\text{CO}) / \Delta V$ (cm$^{-2}$/\kms) & $7 \times 10^{14}$ & $7\times10^{17}$ & $100$ \\

$N(^{13}\text{CO}) / \Delta V$ (cm$^{-2}$/\kms) & $10^{14}$ & $7\times10^{16}$ & $100$ \\

\hline
\end{tabular}

\label{tbl-radexlib-co}
\end{table}

\subsection{Kinetic Temperature and Brightness Temperature}\label{app:tkin-tb}
At frequencies of $\ge346$\,GHz, the condition $h\nu << kT$ is not adequately satisfied for the Rayleigh-Jeans approximation, for the typical range of temperatures we are likely to encounter in the circumstellar medium. Therefore, even assuming that the CO rotational excitation is in LTE, and the line optical depths are greater than unity, the brightness temperature will generally be less than the kinetic temperature for the \jtres~and higher transitions. The plot below shows that for \codos\,\jtres~brightness temperature in the range $\sim(15-90)$\,K, the corresponding minimum value of $T_{kin}$ required is $\sim(22-98)$\,K.

\begin{figure}[hbtp]
\includegraphics[width=16cm]{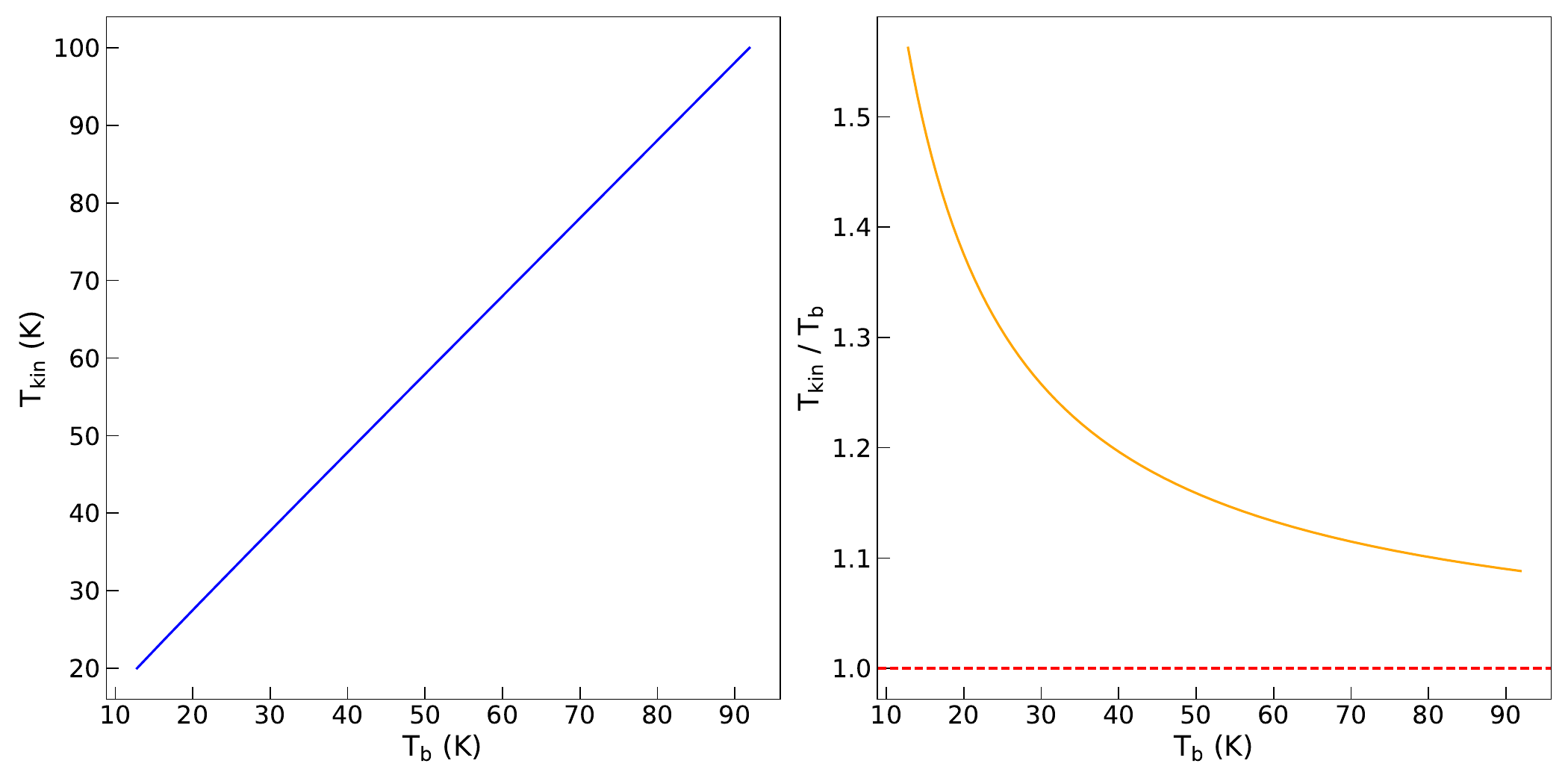}

\caption{Kinetic Temperature (T$_{kin}$) as a function of brightness temperature (T$_b$) for the \codos\,\jtres~line under LTE conditions. Panel $b$ shows the ratio T$_{kin}$/T$_b$ as a function ofT$_b$.}
\label{tkinvstb}
\end{figure}

\subsection{Modeling Strategy: CBN}\label{app:modCBN}
In the CBN, the lower limit to the kinetic temperature (that likely varies from the bright regions in the torus to the fainter regions that lie near the edge of the CBN along its long axis) is set using the constraint that the kinetic temperature has to be larger than the observed brightness temperatures of the collisionally-excited CO lines. But it is important to note the limitations of the brightness temperature as a measure of the excitation temperature ($T_{exc}$) and the minimum kinetic temperature required to produce any given value of $T_{exc}$. At frequencies of $\ge346$\,GHz, the condition $h\nu << kT$ is not adequately satisfied for the Rayleigh-Jeans approximation, for the typical range of temperatures we are likely to encounter in the circumstellar medium. Therefore, even assuming that the CO rotational excitation is in LTE, and the line optical depths are greater than unity, the brightness temperature will be less than the kinetic temperature for the \jtres~and \jsix~lines (see \S\,\ref{app:tkin-tb}).

We find that the \codos\,\jtres~brightness temperature spans the range $\sim(31-67)$\,K in the CBN as seen in the \jtres~emission at systemic velocity; the median value is $39$\,K, and the corresponding minimum value of $T_{kin}$ required is $47$\,K\footnote{we use the \jtres~line instead of the \jsix~line, since the former is brighter than the latter everywhere in IRAS\,06530}. We set the minimum average kinetic temperature in the CBN to $47$\,K, and fit the observed mean \codos\,\jtres~(\codos\,\jsix)~line intensity of the CBN at the systematic velocity $31.9$\,K ($23.5$\,K), conservatively assuming a $\pm10$\,\% error. The results show that $T_{kin}$ remains unconstrained at the upper end: for these high values of $T_{kin}$, the densities are low and the \codos\,\jtres~and \jsix~lines are (severely) subthermally excited. We therefore take into account the results from the NOEMA data, which show that the average brightness temperature (at line-center) of IRAS\,06530 in the \codos\,\jdos~line, extracted from a $2{''}\times2{''}$ aperture, is $T_{^{12}CO[N]}=22.1$\,K. In order to scale this value to our smaller CBN aperture and smaller beam-size, we first convolve our \codos\,\jtres~data to the same resolution as the NOEMA beam (hereafter [ALMA-to-NOEMA]\codos\,\jtres~datacube), and then extract the spectrum from a $2{''}\times2{''}$ aperture -- we find that the brightness temperature (at line-center) in our data is a factor $1.48$ higher than in the latter. We then assume that a similar scaling applies to the \jdos~data, implying that the average \jdos~brightness temperature (at line-center) in the CBN aperture is $T_{^{12}CO[N]}(scl)=32.7$\,K (with an assumed conservative $\pm$15\% uncertainty). Applying this constraint, we find that the maximum value of $T_{kin}$ that fits the CBN \codos\,\jtres~and \codos\,\jsix~mean intensities is $57$\,K; for higher $T_{kin}$ values, the model \jdos~brightness temperature is lower than the lower limit on $T_{^{12}CO[N]}(scl)$, $27.8$\,K.

We then attempt to fit the \cotres\,\jtres~data, which includes the\cotres\,\jtres~mean observed intensity at the systemic velocity of $2.2$\,K, as well as the mean (velocity-integrated) \cotres\,\jtres~intensity of $54.6$\,K-\kms~averaged for the CBN aperture, using models with the same combinations of kinetic temperature, gas density, and molecular masses that fit the \codos~data as described above. An an additional constraint, we use the \cotres\,\jdos~brightness temperature determined by \cite{sun25} at line-center for their NOEMA aperture ($T_{^{13}CO[N]}=0.62$\,K). We multiply this value to the CBN aperture and our small beamsize, by a scale-factor 1.55, that is derived by applying the same procedure as for \codos\,\jdos~above, to the \cotres\,\jtres~data convolved to the same resolution as the NOEMA beam ([ALMA-to-NOEMA]\cotres\,\jtres), resulting in $T_{^{13}CO[N]}(scl)=0.96$\,K. We find that only models with the highest \tkin~value ($57$\,K) result in a \cotres\,\jdos~average intensity that is consistent with the maximum allowed value of $T_{^{13}CO[N]}(scl)$ ($1.15$\,K), assuming a $\pm$20\% uncertainty; for lower \tkin~models, the model \cotres\,\jdos~average intensity is significantly higher than the observed value.

Then, using the resulting model values of $N_{CO} / \delta V$, and the mean (velocity-integrated) \codos\,\jsix~intensity averaged over the CBN\footnote{we use the \codos\,\jsix~flux instead of the \codos\,\jtres~flux because the former does not include any contribution from the ring} of $4.456\times10^2$\,K-\kms, the total projected area of the CBN of $3.64\times10^7$\,au$^2$, and an assumed CO abundance ratio of, CO/H$_2$ = $f_{CO}=4\times10^{-4}$, we find that, with $T_{kin}=57$\,K the fitted molecular mass of the CBN is $M_{H_2} = 2.2\times10^{-2}$\,\ms, and \codos/\cotres$=24$. The average optical depths of the \codos\,\jtres, \codos\,\jsix, and \cotres\,\jtres~lines for the final set of best-fit models are $(0.99-1.5)$, $(1.1-1.6)$, and $(0.044-0,058)$.

\subsection{Modeling Strategy: Torus}\label{app:modtorus}

We assume a minimum average kinetic temperature of $60$\,K in the torus, based on the median brightness temperature of the \codos\,\jtres~line. The observed mean \codos\,\jtres~(\codos\,\jsix)~line intensity of the torus at the systemic velocity is $52$\,K($46$\,K). The resulting model fits show that $T_{kin}$ remains unconstrained at the upper end (as was the case for the CBN). We therefore take into account the results from the NOEMA \codos\,\jdos~data, which show that the average brightness temperature (at systemic velocity) of IRAS\,06530 in the torus region, corrected (upwards) for the smaller beam, is conservatively $T_b$({\codos\,\jdos})$\gtrsim45$\,K, as estimated from Fig.\,5 of \cite{sun25}\footnote{the panel at systemic velocity in this figure shows that the intensity in the torus region is $T_b$(\codos\,\jdos)$\gtrsim0.4$\,Jy\,beam$^{-1}$, i.e., $31.2$\,K for their beam of $0\farcs72\times0\farcs41$. We correct this intensity for our smaller beam by multiplying it with a scale factor $1.45$, which is the ratio of our observed \codos\,\jtres~average torus intensity with our small beam to that extracted from the [ALMA-to-NOEMA]\codos\,\jtres~datacube at systemic velocity}. Applying this constraint, the highest allowed value of $T_{kin}$ is $220$\,K; for higher values of $T_{kin}$, the model \codos\,\jdos~intensity falls progressively below its lowest allowed value of $36$\,K, assuming a $\pm20$\% uncertainty in $T_b$({\codos\,\jdos}).

We then apply the following further constraints: the torus mass is less than that of the CBN, and that the torus' average density is greater than that of the CBN, but by a factor $\le3$; the latter is a conservative choice based on our dust modelng that gives a maximum overdensity factor of $1.9$. This results in $T_{kin}=60-99$\,K and \moldens$=(0.96-14.4)\times10^4$\,\vdensunit. Now, by further requiring that the torus mass is constrained to lie in the range, $(0.52-0.73)\times10^{-2}$\,\ms~(see \S\,\ref{torus_mass}), we find that $T_{kin}=64-85$\,K, and
using the resulting model values of $N_{CO} / \delta V$, and the mean (velocity-integrated) \codos\,\jsix~intensity of $7.03\times10^2$\,K-\kms, averaged over the projected area of the torus $7.137\times10^{6}$\,au$^2$, we establish the final set of \codos~models.

We estimate a \cotres\,\jdos~mean intensity in the torus at the systemic velocity from Fig.\,8 of \cite{sun25} and re-scale it (as for \codos\,\jdos~above), obtaining $2.2$\,K ($\pm$\,20\%). Then, using the physical parameters established above, the mean observed intensity of \cotres\,\jtres~in the torus at the systemic velocity, $4$\,K ($\pm$\,10\%) and that of \cotres\,\jdos, together with the mean (velocity-integrated) \cotres\,\jtres~intensity, $92$\,K-\kms, we derive the \codos/\cotres~abundance ratio. We find that \codos/\cotres$=15-21$. The average optical depths of the \codos\,\jtres, \codos\,\jsix, and \cotres\,\jtres~lines for the final set of best-fit models are $(1.1-1.8)$, $(1.7-2.5)$, and $(0.052-0.073)$.

We derive the \coeksaat/\cotres~abundance ratio in the torus, using (i) the average intensities of the \coeksaat\,\jsix~and \cotres\,\jtres~lines in the torus, and (ii) a grid of RADEX models that spans the appropriate range of physical conditions derived above. Since the \coeksaat\,\jsix~emission is relatively weak, we average it (as well as the \cotres\,\jtres~line emission) over a $\sim18.4$\,\kms~range around the systemic velocity in order to determine its mean intensity in the torus -- $2.0$\,K ($3.6$\,K). This choice of the velocity range is relatively large in order to minimize the intensity uncertainties, while still excluding any contribution from the ring emission. Conservatively assuming a $\pm$10\,\% error in these intensity values, we find that the minimum (maximum) \coeksaat/\cotres~abundance ratio is $0.39$ ($1.0$).

\subsection{Modeling Strategy: Ring}\label{app:modring}
The peak \codos\,\jtres~intensities of R-Shrp or R-Plat together with the upper limit on the \codos\,\jsix, and the ``geometrical" constraint,  do not allow us to set meaningful constraints on the kinetic temperature or density. We therefore assume a kinetic temperature of $T_{kin} \ge 50$\,K, based on the $T_{kin}$ values derived by \cite{schoier2005} for detached shells in C-rich AGB stars ($170\pm100$ in one, and $>60$\,K in 5/7 objects). We find that the allowed column and volume density ranges are now significantly restricted. For R-Shrp, with $R_{N/n} = 2\times10^{16}$\,cm/\kms~(and assuming a $\pm15$\.\% uncertainty), we find: $4.35\times10^{15} \le N_{CO} / \delta V\,({\rm cm}^{-2} / \kms) \le 9.9\times10^{15}$, and $545 \le n_{H_2}\,({\rm cm}^{-3}) \le 1255$). The average optical depth of the \codos\,\jtres~line for the final set of best-fit models is $(0.81-2.0)$.

We integrate the density of R-Shrp (assuming it has a gaussian profile with the same shape as its intensity) over the volume of a shell with radial width equal to $1\times$FWHM, and find that it has a mass of $(0.95-2.2)\times10^{-2}$\,\ms~corresponding to the minimum and maximum values of $N_{CO} / \delta V$.

Using a similar methodology for R-Plat, with $R_{N/n} = 2.1\times10^{16}$ cm/\kms~(and assuming a $\pm15$\.\% uncertainty), we find that the column and volume densities are, respectively, $3.7\times10^{15} \le N_{CO} / \delta V\,({\rm cm}^{-2} / \kms) \le 7.0\times10^{15}$, and $500 \le n_{H_2}\,({\rm cm}^{-3}) \le 830$. The average optical depth of the \codos\,\jtres~line for the final set of best-fit models is $(0.71-1.4)$.

The mass of R-Plat lies in the range $(2.4-4.4)\times10^{-2}$\,\ms, assuming it's density has a gaussian profile with the same shape as its intensity with radial width equal to $1\times$FWHM. The resulting total ring mass is $(3.3-6.6)\times10^{-2}$\,\ms, with R-Plat containing $\sim 70$\,\% of the mass.

\subsection{Modeling Strategy: FAR}\label{app:modfar}
We apply the same strategy for modeling the FAR emission as for the ring, using the peak \codos\,\jtres~and \jsix~intensities of the FAR, together with the ``geometrical" constraint. We assume $\pm$\,10\% (30\%) uncertainy in the peak \codos\,\jtres~(\jsix) intensities. The peak \codos\,\jtres~intensity implies a minimum kinetic temperature of $13.7$\,K. We find that $R_{N/n} = 5.2\times10^{15}$\,cm/\kms~(and assuming a $\pm15$\.\% uncertainty), we find: $1.75\times10^{17} \le N_{CO} / \delta V\,({\rm cm}^{-2} / \kms) \le 3.7\times10^{17}$, and $0.83\times10^5 \le n_{H_2}\,({\rm cm}^{-3}) \le 1.8\times10^5$.

We integrate the density of the FAR (assuming it has a gaussian profile with the same shape as its intensity) over the volume of a shell with radial width equal to the FAR's FWHM, and find that it has a mass of $(0.12-0.25)$\,\ms~corresponding to the minimum and maximum values of $N_{CO} / \delta V$.

\end{document}